\documentclass[]{aa}  

\usepackage{natbib}
\bibpunct{(}{)}{;}{a}{}{,}
\usepackage{graphicx}
\usepackage{revsymb}	
\usepackage{amsmath}	
\usepackage[usenames]{color}	
\usepackage{amssymb}
\usepackage{color}
\usepackage{geometry} 
\usepackage[parfill]{parskip} 
\usepackage{amssymb}
\usepackage{slantsc}
\usepackage{array}
\usepackage{url}
\usepackage{amsfonts}
\usepackage{sidecap}
\usepackage{xcolor}
\usepackage{nicefrac}
\usepackage{epsfig}
\colorlet{rouge}{red!70!darkgray}
\usepackage[varg]{txfonts}
\usepackage{multirow}
\usepackage{caption}
\usepackage{subcaption}
\usepackage[hidelinks]{hyperref}

\begin{document}
   \title{Coralie radial velocity search for companions\\around evolved stars  (CASCADES)}
      \subtitle{II. Seismic masses for three red giants orbited by long-period massive planets. \thanks{Based on observations collected with the Coralie echelle spectrograph on the 1.2-m Euler Swiss telescope at La Silla Observatory, ESO, Chile.}}
\author{G. Buldgen\inst{1} \and G. Ottoni\inst{1} \and C. Pezzotti\inst{1} \and A. Lyttle\inst{2} \and P. Eggenberger\inst{1} \and S. Udry\inst{1} \and D. Ségransan\inst{1} \and A. Miglio\inst{3,4,2} \and M. Mayor\inst{1} \and C. Lovis\inst{1} \and Y. Elsworth\inst{2} \and G.R. Davies \inst{2} \and W.H. Ball\inst{2}}
\institute{Département d'Astronomie, Université de Genève, Chemin Pegasi 51, CH$-$1290 Sauverny, Suisse. \and School of Physics and Astronomy, University of Birmingham, Edgbaston, Birmingham B15 2TT, UK.
\and Dipartimento di Fisica e Astronomia, Universit{\`a} degli Studi di Bologna, Via Gobetti 93/2, I-40129 Bologna, Italy \and INAF -- Astrophysics and Space Science Observatory Bologna, Via Gobetti 93/3, I-40129 Bologna, Italy.}

\date{December 7th, 2020}

\abstract{The advent of asteroseismology as the golden path to precisely characterize single stars naturally led to synergies with the field of exoplanetology. Today, the precise determination of stellar masses, radii and ages for exoplanet-host stars is a driving force in the development of dedicated software and techniques to achieve this goal. However, as various approaches exist, it is clear that they all have advantages and inconveniences and that there is a trade-off between accuracy, efficiency, and robustness of the techniques.}
{We aim to compare and discuss various modelling techniques for exoplanet-host red giant stars for which TESS data are available. The results of the seismic modelling are then used to study the dynamical evolution and atmospheric evaporation of the planetary systems.}{We study, in detail, the robustness, accuracy and precision of various seismic modelling techniques when applied to four exoplanet-host red giants observed by TESS. We discuss the use of global seismic indexes, the use of individual radial frequencies and that of non-radial oscillations. In each case, we discuss the advantages and inconveniences of the modelling technique.}{We determine precise and accurate masses of exoplanet-host red giant stars orbited by long-period Jupiter-like planets using various modelling techniques. For each target, we also provide a model-independent estimate of the mass from a mean density inversion combined with radii values from \textit{Gaia} and spectroscopic data. We show that no engulfment or migration is observed for these targets, even if their evolution is extended beyond their estimated seismic ages up the red giant branch.}{}

\keywords{Stars: planetary systems - Stars: interiors - Stars: fundamental parameters - Asteroseismology - Planet–star interactions - Stars: individual - HD 22532, HD 64121, HD 69123}

\maketitle
\section{Introduction}

With the recent breakthroughs in the field of asteroseismology, as a result of the so-called `space-based photometry revolution' that was started by the CoRoT \citep{Baglin} and \textit{Kepler} \citep{Borucki} satellites and is still in full swing with TESS \citep{Ricker2015}, this young field of stellar physics has become a natural complement to the field of exoplanetology. Indeed, the need for precise and accurate stellar parameters remains a key issue for stellar modellers and a central aspect for the preparation of future missions such as PLATO \citep{Rauer}. This led to numerous synergies between exoplanetology and asteroseismology, illustrating the common interests of both fields in providing accurate depictions of exoplanet-host stars \citep[see e.g.][amongst others]{JCDKOI2010, Batalha2011, Huber2013,Huber2013Science,Davies2015,Silva2015,Lundkvist2016,Davies2016,Huber2018,Campante2018}. Beyond the need for accurate and precise planetary masses and radii, a good depiction of the host star is also required to properly understand the habitability, formation history, and dynamical evolution of planetary systems \citep{Huber2013Science,Farr2018}. Indeed, having access to temporal information by using stellar models also allows to put the current state of observed planetary systems in a `historical' perspective, permitting one to constrain additional physical phenomena such as the effects of tides and atmospheric evaporation on the architecture of exoplanetary systems. 

In this paper, we demonstrate the importance of thorough seismic modelling for exoplanet-host red giant branch (RGB) stars. We show that going beyond the use of the standard seismic scaling relations is of primary importance to determine robust and accurate stellar parameters. We applied dedicated seismic modelling techniques to three red giants observed by TESS, for which long-period massive planets have been detected by the CASCADES survey (Ottoni et al. submitted), namely $\rm{HD22532}/\rm{TIC200841704}$, $\rm{HD64121}/\rm{TIC264770836}$, and $\rm{HD69123}/\rm{TIC146264536}$ that have been detected using radial velocity measurements with the CORALIE echelle spectrograph. All of these planets are long period Jupiter-like planets with periods of $872$, $623$, and $1193$ days, for which we could determine their masses to be between $2$ and $3$ Jupiter masses (Ottoni et al. submitted).

We start in Sect. \ref{AsteroObs} by describing the peakbagging procedure used to determine the individual pulsation frequencies for all targets. We focus on the analysis of radial $(\ell=0)$ and quadrupole $(\ell=2)$ modes which are used afterwards in the modelling. In Section \ref{SecModelling}, we detail the various procedures used to determine the fundamental properties of each star. We first illustrate in Sect. \ref{SecGlobalIndex} the use of the seismic scaling relations, discussing the impact of the various corrections proposed in the literature and the potential issues related to the robustness and accuracy of these techniques. 

In Sects. \ref{SecIndFreq} and \ref{SecFreqInv}, we present the results of seismic modelling using individual frequencies and frequency differences, coupled to stellar evolutionary models. We compare the results of this dedicated modelling of each target to the various results of the scaling relations and discuss the implications of our tests for the requirements of precise stellar parameters for the purpose of exoplanetology. Finally, in Sect. \ref{SecOrbital}, we discuss the dynamical evolution of the planetary systems for each target under the effects of both dynamical and equilibrium tides in a fully coupled way with the evolution of the host star.
  
\section{Determination of oscillation parameters and frequencies}\label{AsteroObs}
We measured the observed asteroseismic oscillation mode frequencies, $\nu_{n, \ell}$, for the three targets using TESS photometric time series. We first constructed power spectra from the observed photometric flux. Using the peakbagging package \texttt{PBjam}\footnote{See \url{https://github.com/grd349/PBjam}} \citep{Nielsen2020}, we then determined the locations of radial and quadrupolar oscillation modes. Some evidence of dipolar modes could be seen in the spectra; however, as a result of the short duration of the observations, the frequency resolution could not allow us to provide a clear and unambiguous identification of these modes.

\begin{table}
    \begin{center}
        \caption{TESS sectors available for each target.}
        \label{tab:sectors}
        \begin{tabular}{c|c|c}
            \hline\hline
            \textbf{HD ID} & \textbf{TIC ID} & TESS Sectors \\
            \hline
            HD69123 & TIC146264536 & 7, 8 \\
            HD22532 & TIC200841704 & 3, 4 \\
            HD64121 & TIC264770836 & 8, 9 \\
            \hline
        \end{tabular}
    \end{center}
\end{table}

Using the \texttt{lightkurve} package \citep{lightkurvecollaborationLightkurveKeplerTESS-2018} with \texttt{astropy} \citep{AstropyCollaboration.Robitaille.ea2013,AstropyCollaboration.Price-Whelan.ea2018} and \texttt{astroquery} \citep{Ginsburg.Sipocz.ea2019}, we constructed the power spectra as follows. We downloaded TESS light curves from the Mikulski Archive for Space Telescopes (MAST) for the available sectors given in Table~\ref{tab:sectors}. For each star, we stitched the pre-search data conditioning simple aperture photometry \citep[PDCSAP,][]{Stumpe.Smith.ea2012,Smith.Stumpe.ea2012} flux from each sector together. We discarded $5$-$\sigma$ outliers and removed low-frequency trends using a Savitzky-Golay filter. We then determined the power spectrum for each star using the Lomb-Scargle method \citep{Lomb1976,Scargle1982} and divided the power by an estimate of the background to determine the signal-to-noise ratio (S/N; where an S/N of 1 indicates the absence of any signal).

\begin{table*}
    \begin{center}
    \caption{Global stellar properties used as inputs for the PBjam peakbagging pipeline.}
    \label{tab:seismo_input}
    \begin{tabular}{c|c|c|c}
    \hline \hline
    \textbf{Input} & HD22532 &  HD64121 & HD69123 \\
    \hline
    $\Delta\nu\,(\mu\mathrm{Hz})$ & $10.68\pm0.07$ & $11.33\pm0.09$ & $7.32\pm0.07$ \\
    $\nu_{\mathrm{max}}\,(\mu\mathrm{Hz})$ & $129.38\pm2.12$ & $138.51\pm1.32$ & $88.58\pm1.54$\\
    $T_{\mathrm{eff}}\,(\mathrm{K})$ & $5067\pm70$ & $4980\pm70$ & $4787\pm70$ \\
    $G_{\mathrm{BP}} - G_{\mathrm{RP}}\,(\mathrm{dex})$ & $1.09\pm0.01$ & $1.08\pm0.01$ & $1.18\pm0.01$  \\
    \hline
    \end{tabular}
    \end{center}
\end{table*}
        
We used the \texttt{PBjam} package to measure the observed radial, $\nu_{n,\ell=0}$, and quadrupolar, $\nu_{n,\ell=2}$, oscillation modes for each target. Initially, we performed the mode identification based on the methods of \citet{daviesAsteroseismologyRedGiants-2016} and a prior probability distribution constructed from thousands of stars already analysed using \texttt{PBjam}. The values and uncertainties of input parameters used to select stars from the prior are given in Table~\ref{tab:seismo_input}. The input frequency at maximum power, $\nu_{\mathrm{max}}$, and the large frequency separation, $\Delta\nu$ were determined using the layered approach of \citet{Elsworth_2020}, combining for $\Delta\nu$ the power spectrum of the power spectrum approach \citep{Hekker2010}, the universal pattern method \citep{Mosser2011}, and two other criteria for the mode properties \citep[See][ for additional details]{Elsworth_2020}. We also adopted the input effective temperatures, $T_{\mathrm{eff}}$, and colours, $G_{\mathrm{BP}} - G_{\mathrm{RP}}$, from \textit{Gaia} DR2 \citep{GaiaCollaboration.Prusti.ea2016,gaiacollaborationGaiaDataRelease-2018}. The inputs primarily determined the window in which we selected stars from the prior for subsequent mode identification. As a result, the inputs have very little effect on the final peakbagging step.

We identified initial mode locations, $\nu_{n,\ell}^{\,\prime}$, by fitting the asymptotic relation \citep{2013A&A...550A.126M} to 9 radial orders using Bayes' Theorem, $P(\boldsymbol{\theta}|D) \propto P(\boldsymbol{\theta}) P(D|\boldsymbol{\theta})$. The likelihood of the S/N data, $D$, given the model parameters, $\boldsymbol{\theta}$, is given by the term $P(D|\boldsymbol{\theta})$. The prior distribution of the model, $P(\boldsymbol{\theta})$, was obtained using a kernel density estimate for the population selected in the previous step. We sampled the posterior, $P(\boldsymbol{\theta}|D)$, using the \texttt{emcee} package \citep{foreman-mackeyEmceeMCMCHammer-2013}.
  
\begin{table*}
    \begin{center}
    \caption{Individual observed radial $(\ell=0)$ and quadrupolar $(\ell=2)$ oscillation modes, with their statistical uncertainties for each star.}
    \label{tab:seismo_output}
    \begin{tabular}{cc|cc|cc}
        \hline \hline
        \multicolumn{2}{c|}{HD69123} & \multicolumn{2}{c|}{HD22532} & \multicolumn{2}{c}{HD64121}\\
        \hline
         $\nu_{n,0}\,(\mu\mathrm{Hz})$ & $\nu_{n,2}\,(\mu\mathrm{Hz})$ & $\nu_{n,0}\,(\mu\mathrm{Hz})$ & $\nu_{n,2}\,(\mu\mathrm{Hz})$ & $\nu_{n,0}\,(\mu\mathrm{Hz})$ & $\nu_{n,2}\,(\mu\mathrm{Hz})$ \\
        \hline
         $66.90\pm0.12$ &                           --- &                           --- &                $96.53\pm0.25$ &               $104.64\pm0.30$ &                           --- \\
         $74.00\pm0.15$ &                           --- &               $108.23\pm0.14$ &               $107.07\pm0.24$ &               $115.99\pm0.27$ &               $114.22\pm0.29$ \\
         $81.29\pm0.15$ &                $80.41\pm0.18$ &               $118.95\pm0.10$ &                           --- &               $127.02\pm0.10$ &               $125.74\pm0.29$ \\
                $88.43\pm0.20$ &                $87.51\pm0.13$ &               $129.45\pm0.10$ &               $128.15\pm0.24$ &               $138.44\pm0.23$ &               $136.93\pm0.24$ \\
         $95.82\pm0.05$ &                $94.97\pm0.09$ &               $140.07\pm0.11$ &                           --- &               $149.67\pm0.11$ &                           --- \\
         $103.36\pm0.10$ &               $102.48\pm0.20$ &               $150.73\pm0.18$ &               $149.46\pm0.15$ &               $161.06\pm0.14$ &                           --- \\
         $110.80\pm0.16$ &               $109.83\pm0.17$ &                           --- &                           --- &               $172.53\pm0.32$ &               $171.12\pm0.30$ \\
        $118.43\pm0.18$ &                           --- &               $172.31\pm0.30$ &                           --- &                           --- &                           --- \\
       \hline
    \end{tabular}
    \end{center}
\end{table*}

Finally, we fitted a Lorentzian profile to each mode individually by sampling its posterior distribution using the Bayesian package \texttt{PyMC3} \citep{Salvatier.Wiecki.ea2016}. We used the initial mode identification as a prior on the Lorentzian centers, such that $\nu_{n,\ell} \sim \mathcal{N}(\nu_{n,\ell}^{\,\prime}, 0.03\Delta\nu)$, where $\mathcal{N}(\mu, \sigma)$ is a normal distribution with mean $\mu$ and standard deviation $\sigma$. All of the other parameters from the previous steps were relaxed. The resulting radial and quadrupolar mode locations are given in Table~ \ref{tab:seismo_output} and marked in Fig.~\ref{fig:seismo_model}. We also provide the power spectrum for each star in Appendix \ref{SecAddPlots}. We discarded results where the uncertainty over the prior standard deviation, $0.03\Delta\nu$, was greater than $95\%$, indicating uninformative data. 




\begin{figure}
    \centering
    \begin{subfigure}[t]{0.47\textwidth}
        \centering
        \includegraphics[width=1.0\textwidth]{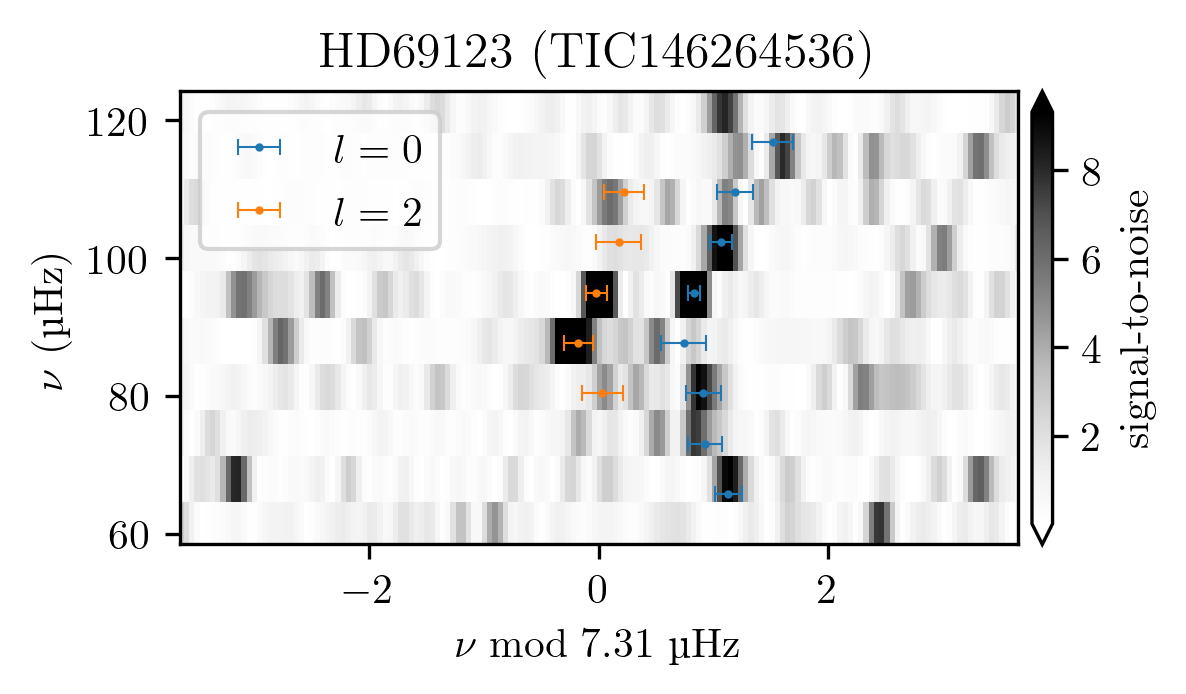}
    \end{subfigure}\\
    ~
    \begin{subfigure}[t]{0.47\textwidth}
        \centering
        \includegraphics[width=1.0\textwidth]{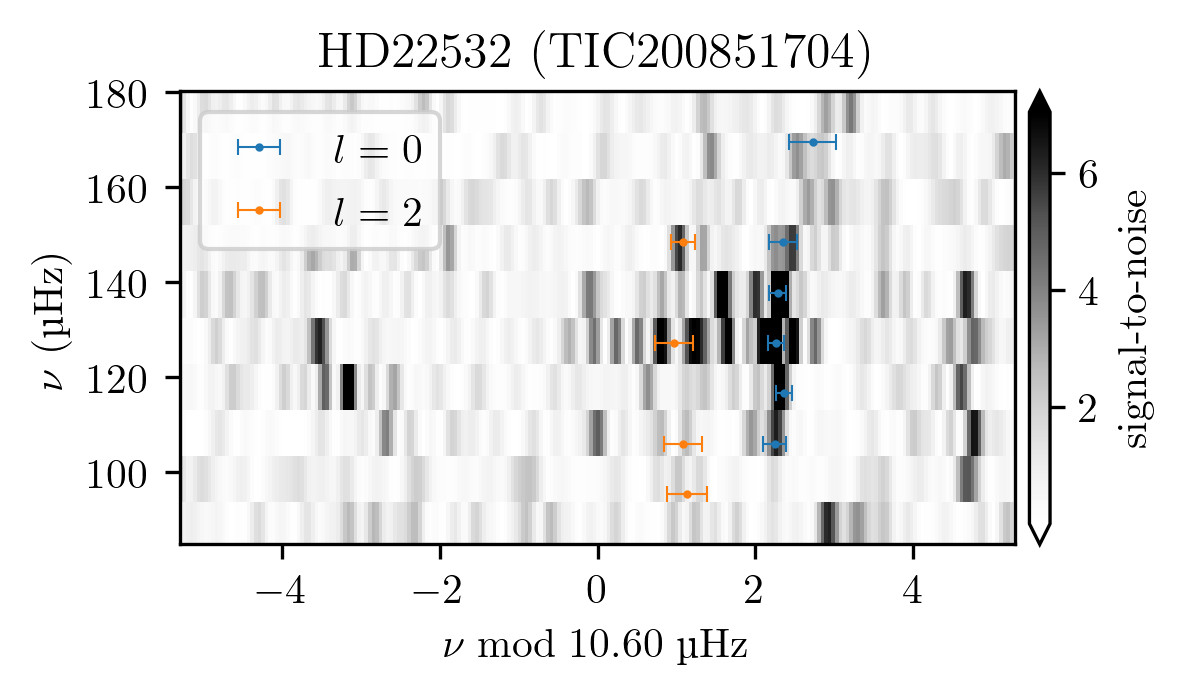}
    \end{subfigure}\\
    ~
    \begin{subfigure}[t]{0.47\textwidth}
        \centering
        \includegraphics[width=1.0\textwidth]{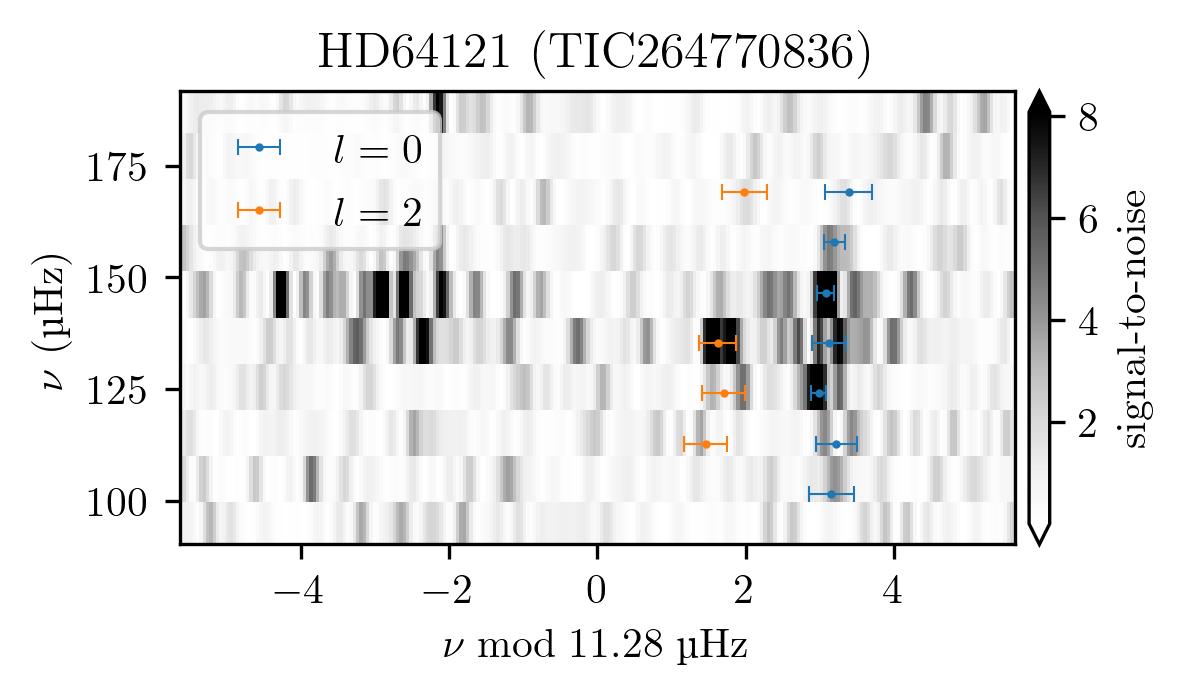}
    \end{subfigure}
    \caption{Observational Echelle diagram for each star. Locations of the radial ($\ell=0$) and quadrupolar ($\ell=2$) oscillation modes are given by blue and orange circles, respectively. The S/N scale is indicated in greyscale on the right-hand side of the plots.}
    \label{fig:seismo_model}
\end{figure}
\section{Seismic modelling}\label{SecModelling}

In this section, we carry out a detailed modelling of all three targets adding at each step a level of refinement in the procedure, discussing the consistency between the different approaches and their intrinsic limitations. First, we start with seismic scaling relations of global seismic indexes in Sect. \ref{SecGlobalIndex}. In Sect. \ref{SecIndFreq}, we carry out the modelling of each target using individual radial frequencies, we then discuss the precision, consistency and limitations of these results with previous approaches. Finally, we add one additional refinement in Sect. \ref{SecFreqInv} to the modelling procedure by carrying out an inversion of the mean density combining it in a fit of the small frequency separations in a Levenberg-Marquardt algorithm, computing the evolutionary models individually. 

We start by presenting the classical constraints available for each target in Table~\ref{tabClassicConst}. We note that the evolutionary status of our targets is unambiguously determined by their global seismic parameters. Such ambiguity could have been an issue for the mean density inversions. Indeed, as demonstrated in \citet{Buldgen2019}, the mean-density inversion can be strongly biased if the evolutionary status of an RGB target is unknown, especially in the case of low-mass clump stars that can be mistaken for high-mass first-ascent RGB stars. We also mention that in the seismic modelling, we use an uncertainty of $70$K on the effective temperature to avoid overfitting, although this has a negligible impact on the final solution as the seismic data are, in comparison, much more precise.

\begin{table*}[t]
\caption{Classical constraints for the targets.}
\label{tabClassicConst}
  \centering
\begin{tabular}{r | c | c | c | c }
\hline \hline
\textbf{Identifiers} & $\rm{HD22532}$ & $\rm{HD64121}$ & $\rm{HD69123}$ \\ \hline
$T_{\rm{eff}}$ $\rm{(K)}$ &$5038\pm24$&$5078\pm22$&$4842\pm41$\\
$\rm{L}$ $\rm{(L_{\odot})}$ &$18.80\pm0.33$&$17.70\pm0.30$&$29.51\pm0.57$\\
$\left[ \rm{Fe}/\rm{H} \right]$ $\left( \rm{dex}\right)$ &$-0.21\pm0.02$&$-0.19\pm0.02$&$0.05\pm0.03$\\
$\rm{\log g}$ $\left( \rm{dex}\right)$ &$3.09\pm0.07$& $2.91\pm0.16$&$2.86\pm0.11$\\
\hline
\end{tabular} \\
\small{\textit{Note:} See Ottoni et al. (submitted) for the discussion on the determination of the non-seismic parameters.}
\end{table*}

\subsection{Scaling relations and global indexes}\label{SecGlobalIndex}

Global seismic indexes have the advantage of providing a quick estimate of the global parameters for solar-like oscillators at a very low numerical cost, making them very useful for the studies of large samples of stars, or studies where the data quality does not allow for a more detailed modelling procedure. The seismic relations were originally presented by \citet{Brown1991} and \citet{Kjeldsen1995}, who originally aimed at predicting the pulsations properties of observed stars, although the link between the large frequency separation and the mean density has been known since \citet{Vandakurov1967}. The original scaling relations were thus written as functions of the observed large frequency separation, $\Delta \nu$, and the frequency of maximum power, $\nu_{\rm{max}}$, from a simple scaling law with the solar reference values.

More recently, the relations were re-arranged and used with the aim of estimating stellar fundamental parameters, such as the mass and radius \citep{Stello2008, Kallinger2010}, and they have since been widely used. However, while their success and usefulness have been recognised in the community, the direct use of the scaling relations to determine mass and radii estimates have come under some criticism, especially the scaling relation between the large frequency separation and the mean density. Indeed, comparisons between seismic and `dynamical' masses for red giant eclipsing binaries have shown the limitations of the use of the scaling relations \citep{Gaulme2016, Brogaard2018,Benbakoura2021}.

A key point mentioned by \citet{Brogaard2018} is the sensitivity of the results for the mass and radius with the applied corrections to relations, especially the $\Delta \nu$ relation. Indeed, some calibrated corrections have been proposed in the literature and applied in practical cases \citep[see e.g.][for some examples]{White2011, Sharma2016, Rodrigues2017, Yu2018, Kallinger2018} while other authors have even discussed the actual value of the large frequency separation to be used when applying them \citep{Mosser2013}. This implies that while formally simple, the scaling relations are also tweaked by correction factors to make them more robust, but these corrections significantly influence the results at the level of precision expected from seismic analyses. Similar conclusions on the sensitivity of the relations to their corrections are reached when comparing seismic parallaxes to those of \textit{Gaia} \citep{Hall2019, Khan2019, Zinn2019}. Indeed, the corrections for the $\Delta \nu$ and $\nu_{\rm{max}}$ relations defined in the literature can actually lead to variations in the determined masses and radii by more than $1\sigma$.
In this section, we illustrate and discuss this lack of robustness of the scaling relations in the context of the requirements of precise mass and radii determinations for the purpose of in-depth studies of exoplanetary systems. The seismic scaling relations for the mass and radius determinations are defined as

\begin{align}
\frac{\rm{M}}{\rm{M_{\odot}}} & \approx \left( f_{\nu_{\rm{max}}}\frac{\nu_{\rm{max}}}{\nu_{\rm{max},\odot}}\right)^{3}\left(f_{\Delta \nu}\frac{\Delta \nu}{\Delta \nu_{\odot}}\right)^{-4} \left(\frac{T_{\rm{eff}}}{T_{\rm{eff},\odot}}\right)^{3/2}, \label{eq:ScalingrelMass}\\
\frac{\rm{R}}{\rm{R_{\odot}}} & \approx \left( f_{\nu_{\rm{max}}}\frac{\nu_{\rm{max}}}{\nu_{\rm{max},\odot}}\right)\left(f_{\Delta \nu}\frac{\Delta \nu}{\Delta \nu_{\odot}}\right)^{-2} \left(\frac{T_{\rm{eff}}}{T_{\rm{eff},\odot}}\right)^{1/2}, \label{eq:ScalingrelRad}
\end{align}

where we have included correction factors $f_{\nu_{\rm{max}}}$ and $f_{\Delta \nu}$ that are commonly used and $\rm{M_{\odot}}$, $\rm{R_{\odot}}$, $\Delta \nu_{\odot}$ and $\nu_{\rm{max},\odot}$ are the solar mass, radius, average large frequency separation and frequency of maximum power, respectively. The solar reference values taken here are those of \citet{Huber2011}, namely $\Delta \nu_{\odot}=135.1 \mu\rm{Hz}$ and  $\nu_{\rm{max},\odot}=3090\mu\rm{Hz}$. In practice, these corrections are often derived from grids of stellar models. Some authors have also advocated for additional terms in the scaling relations \citep{Viani2017} or even non-linear generalisations \citep{Kallinger2018}. Here we focus on the usual form of the scaling relations and compare them to more sophisticated modelling techniques.  Recently, \citet{Li2021} have also investigated the intrinsic scatter of the scaling relations. They make the hypothesis that they provide reference values for stellar parameters and they used the red giant branch bump and the zero age core helium buring stage to measure the scatter in their stellar population. They find a limited scatter around these features when comparing them to their \textsl{Galaxia} synthetic population model \cite{Sharma2011} and point towards the difficulty to reduce the scatter by correcting the $\delta \nu$ scaling relation.

Equations \ref{eq:ScalingrelMass} and \ref{eq:ScalingrelRad} have been widely used in the field of asteroseismology to determine masses and radii of thousands of stars simultaneously. However, a strong weakness resides in the hypothesis of homology between the Sun and the observed star. Local approaches of the problem can also be applied, such as illustrated in \citet{ReeseDens} where the scaling relation between $\Delta \nu$ and the mean density is treated as a variational formula that is applied between a given reference model and the observed star. This local method was, however, found to be suboptimal when compared to the SOLA method, as shown by \citet{ReeseDens}, \citet{Buldgen}, and \citet{Buldgen2019}. Another way of using global seismic parameters consists in fitting the observed values to those given by theoretical models, as is done in the PARAM software \citep{Rodrigues2017}. This approach gets rid of the scaling with respect to the Sun, and it is likely preferable for the direct use of the scaling relations when only global seismic indexes are determined.

\begin{figure}
	\centering
		\includegraphics[width=8cm]{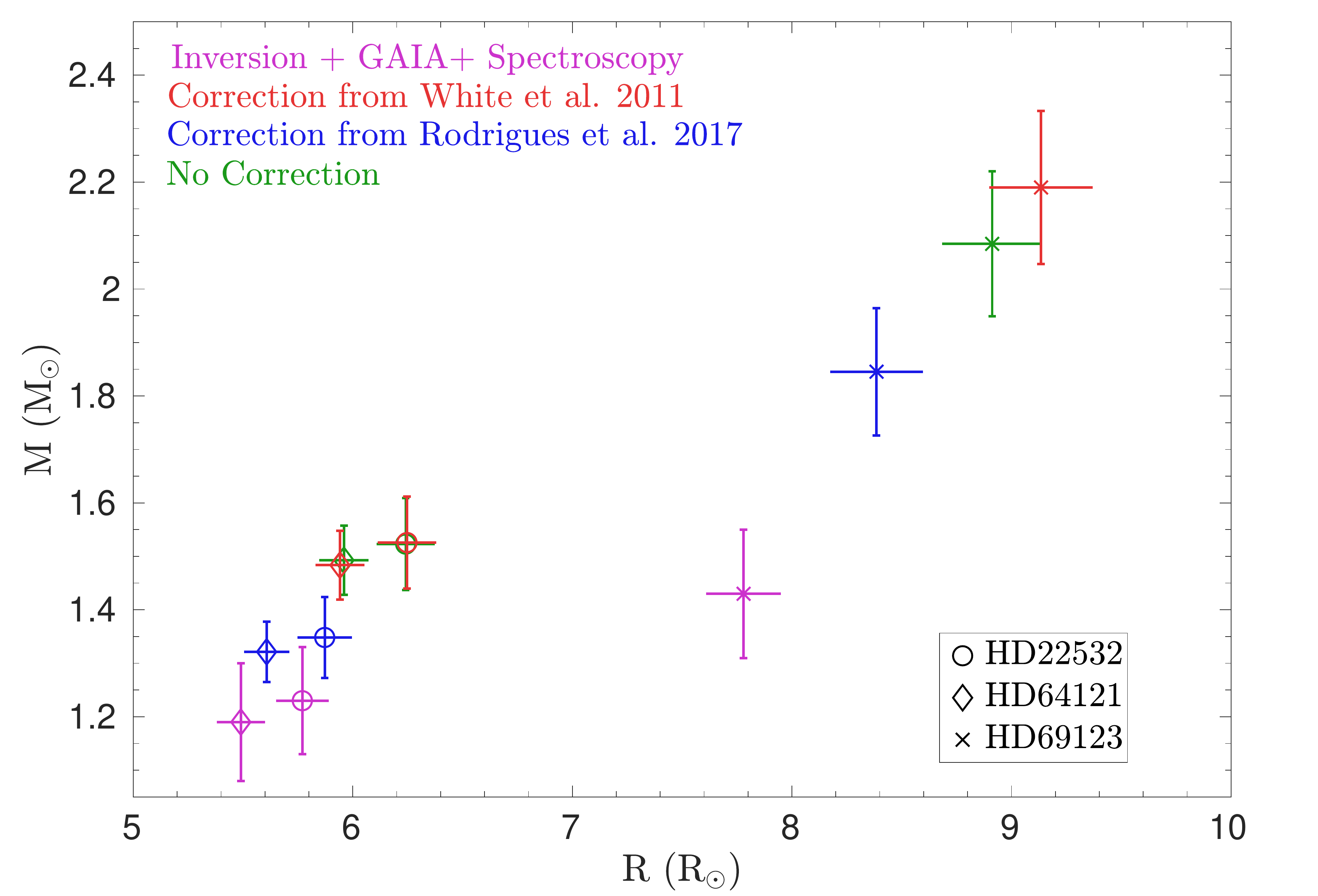}
	\caption{Mass and radii determinations from the scaling relations using various corrections found in the literature as compared to the values determined from the inversion procedure and the radii determined using \textit{Gaia} and spectroscopic constraints shown in Table \ref{tabMassModInd}.}
		\label{FigScaling}
\end{figure}

In Fig. \ref{FigScaling}, we illustrate the results of the application of the `raw' scaling relations, namely fixing $f_{\Delta \nu}$ and $f_{\nu_{\rm{max}}}$ to $1$, as well as the results obtained with two corrections found in the literature\footnote{We note that for HD22532, the correction from \citet{White2011} is almost zero and thus the results are similar to the `raw' scaling relations.}. The first issue we note is that depending on the applied corrections, the results can be significantly different by $1\sigma$ or more. For example, the correction derived from model grids by \citet{Rodrigues2017} leads to a systematic decrease in the estimated mass and radius. While this leads to mass determinations in better agreement with those from detailed modelling (see Sec. \ref{SecIndFreq}) for all stars, the fact that it may lead to variations of more than $1\sigma$ is an indicator that one should use the scaling relations with care. As for the correction based on $\rm{T_{eff}}$ of \citet{White2011}, the changes are of much smaller amplitude for our sample in this case. Overall, the significant differences between the various corrections confirms the results of \citet{Gaulme2016} and \citet{Brogaard2018}. Indeed, in comparing the values for the masses and radii to those determined from seismic inversions with \textit{Gaia} and spectroscopy in table \ref{tabMassModInd}, we can see that despite the corrections, the accuracy is still insufficient for HD69123. Similar conclusions were reached by \citet{Brogaard2018} when analysing the behaviour of the scaling relations for eclipsing binaries in detail.

From this test, it appears that final values of the seismic masses and radii determined from the scaling relations significantly depends on the underlying correction factor $f_{\nu_{\rm{max}}}$ and $f_{\Delta \nu}$ introduced in the scaling relations for the mass and radius, which is unsuitable for the detailed modelling we wish to do here to be able to follow the evolution of the planetary system. Indeed, an accurate and reliable mass determination is required to properly follow the various evolutionary phases that will influence the dynamical evolution of the system as well as the evaporation of the planetary atmospheres. To that end, a detailed modelling seems more suitable as it takes more seismic and non-seismic constraints into account, providing a more reliable solution. Such an example of a more thorough analysis is carried out in the following sections. 

\subsection{Individual frequencies modelling}\label{SecIndFreq}

The modelling using individual frequencies can be carried out using various approaches. In what follows, we make use of the Asteroseismic Inference on a Massive Scale software \citep[AIMS,][]{Rendle2019, Montalban2020} to model the targets and compare the results with the ones from the scaling relations. 

For the purpose of these computations, two separate grids of models were computed. One grid is dedicated to $\rm{HD69123}$ and another grid to $\rm{HD22532}$ and $\rm{HD64121}$. The grid parameters are given in Table~\ref{tabGridProp}\footnote{Fits were also computed with the original grid of \citet{Rendle2019} for the sake of completeness but are not presented here, the results were consistent within $2\sigma$ but differences were seen in $T_{\rm{eff}}$ and $\rm{L}$, as a result of the different boundary conditions of the models of \citet{Rendle2019} that assumed an Eddington $T({\tau})$ relation.}. The grids were computed with the Liège stellar evolution code \citep[CLES,][]{ScuflaireCles} and the adiabatic frequencies were computed with the Liège stellar oscillation code \citep[LOSC,][]{ScuflaireOsc}

\begin{table*}[t]
\caption{Properties of the AIMS stellar evolution models grids.}
\label{tabGridProp}
  \centering
\begin{tabular}{r | c | c }
\hline \hline
\textbf{Parameters}&\textbf{Solar $Z$ grid}& \textbf{Low $Z$ grid} \\ \hline
Mass $\left( M_{\odot} \right)$&$\left[1.00-2.2\right]$ $(0.02\; \rm{step})$ &$\left[1.10-1.90\right]$ $(0.02\; \rm{step})$ $(0.02\; \rm{step})$ \\
$X_{0}$ &$\left[0.68, 0.72 \right]$ $(0.01\; \rm{step})$&$\left[0.71, 0.75 \right]$ $(0.01\; \rm{step})$\\
$Z_{0}$&$\left[ 0.010, 0.040 \right]$ $(0.001\; \rm{step})$&$\left[ 0.006, 0.010 \right]$ $(0.001\; \rm{step})$\\
$\alpha_{\rm{MLT}}$&$2.03$&$2.03$\\
$\nu_{\rm{max}}$ cutoff $\left( \mu \rm{Hz}\right)$&$40$&$60$\\
\hline
\end{tabular}
\end{table*}

The solar mixture used for the grid is that of \citet{AGSS09}, the OPAL opacities \citep{OPAL} and the FreeEOS equation of state were used \citep{Irwin}. In all grids, a solar-calibrated value of the mixing-length parameter fixed at $2.03$ has been used. The atmosphere model used is Model C of \citet{Vernazza}. Microscopic diffusion was not considered in the models and a core overshooting value of $0.15\rm{H_{P}}$, with $\rm{H_{P}}=\left(\frac{dr}{d \ln P}\right)$ being the pressure scale height, was used as well as an envelope overshooting value of $0.15\rm{H_{P}}$. The overshooting regions are considered fully mixed in all cases and the temperature gradient was fixed to the adiabatic gradient in the case of core overshooting and to the radiative gradient in the case of envelope overshooting. The inclusion of core overshooting, although here quite crude, was motivated by its requirement to reproduce seismic observations by \citet{Deheuvels2016}, \citet{Bossini2015}, and \citet{Bossini2017} as well as for eclipsing binaries \citep{Claret2016,Claret2018,Claret2019}, whereas envelope overshooting has been found to be required to reproduce the position of the RGB-Bump in Kepler data \citep{Khan2018}. We also mention here that neglecting microscopic diffusion in the models leads to biases in our determinations of fundamental parameters, especially the age determination that is dominated by the duration of the MS phase. In addition, uncertainties regarding the properties of core overshooting dominate the uncertainties on the ages we report\footnote{This remains true at all stages of our modelling procedure, whatever the seismic analysis performed.}, as all stars exhibit a convective core on the main sequence.  

\begin{figure*}
	\centering
		\includegraphics[width=15cm]{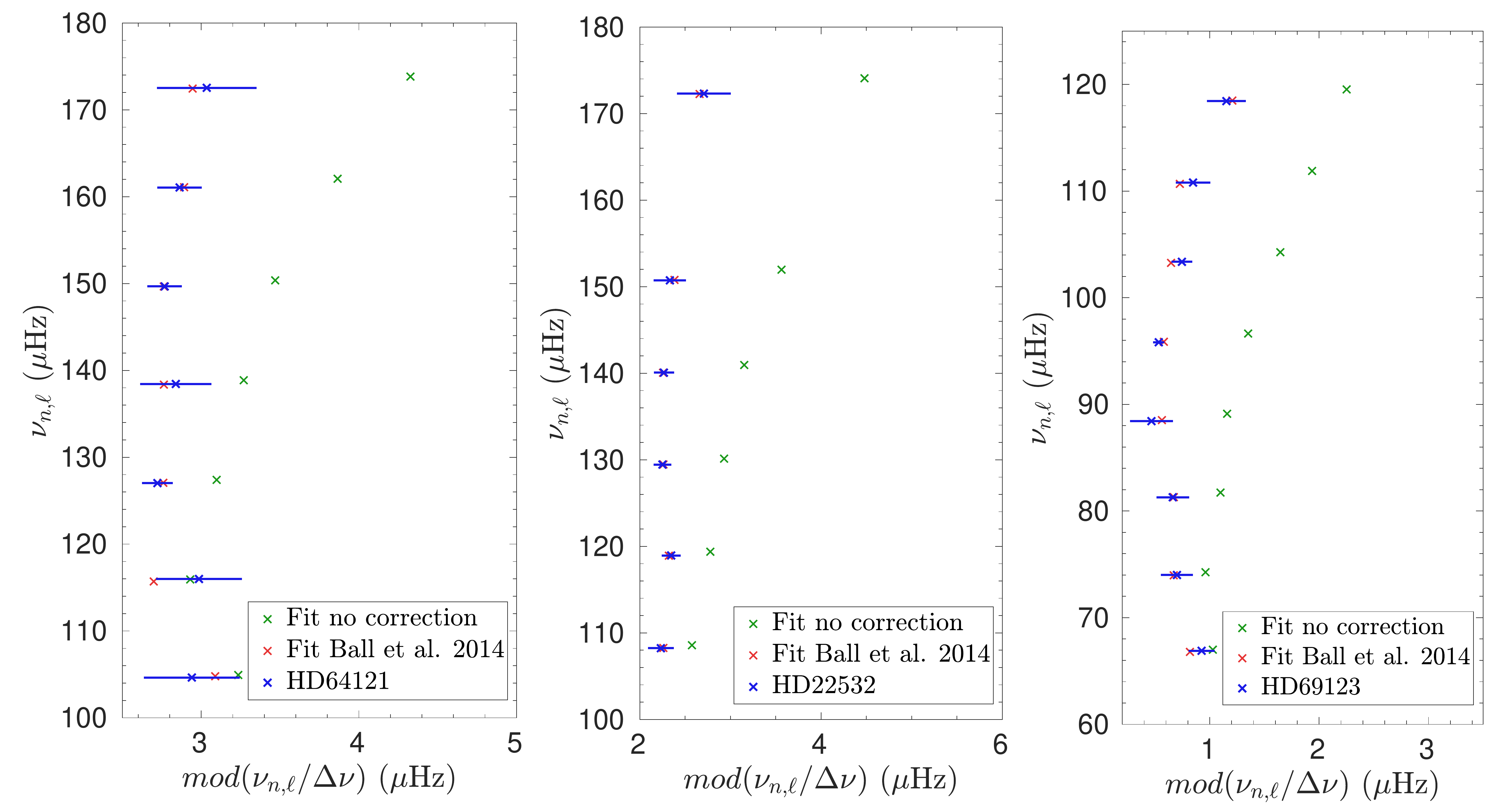}
	\caption{Echelle diagram illustrating the agreement between theoretical and observed radial frequencies for the AIMS solutions using the grids of table \ref{tabGridProp}.}
		\label{FigEchAIMS}
\end{figure*}

\begin{figure*}
\begin{flushleft}
	\begin{minipage}{\textwidth}
  	\includegraphics[trim=70 25 67 95, clip, width=0.325\linewidth]{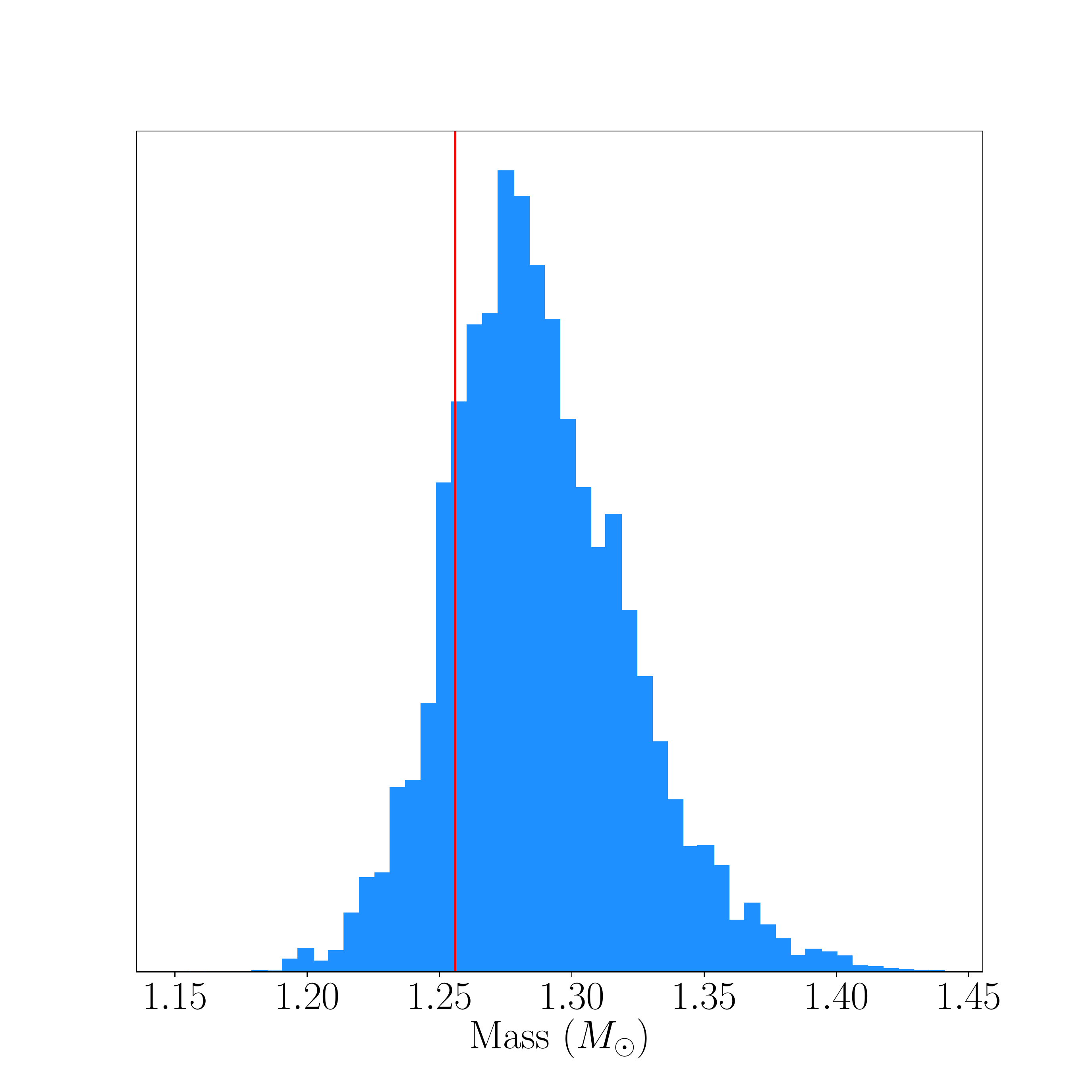} \includegraphics[trim=70 25 70 95, clip, width=0.325\linewidth]{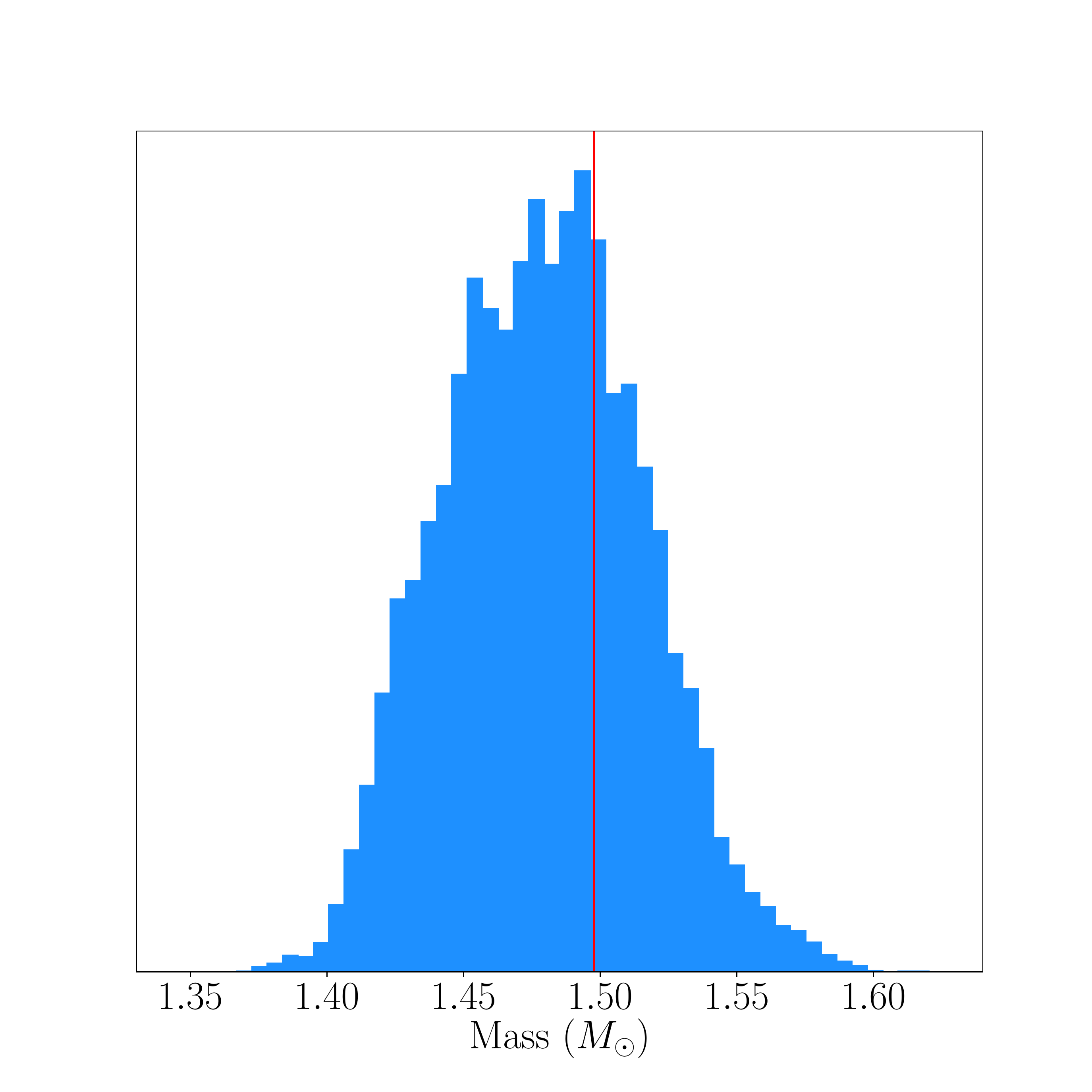}
 	\includegraphics[trim=70 25 70 95, clip, width=0.325\linewidth]{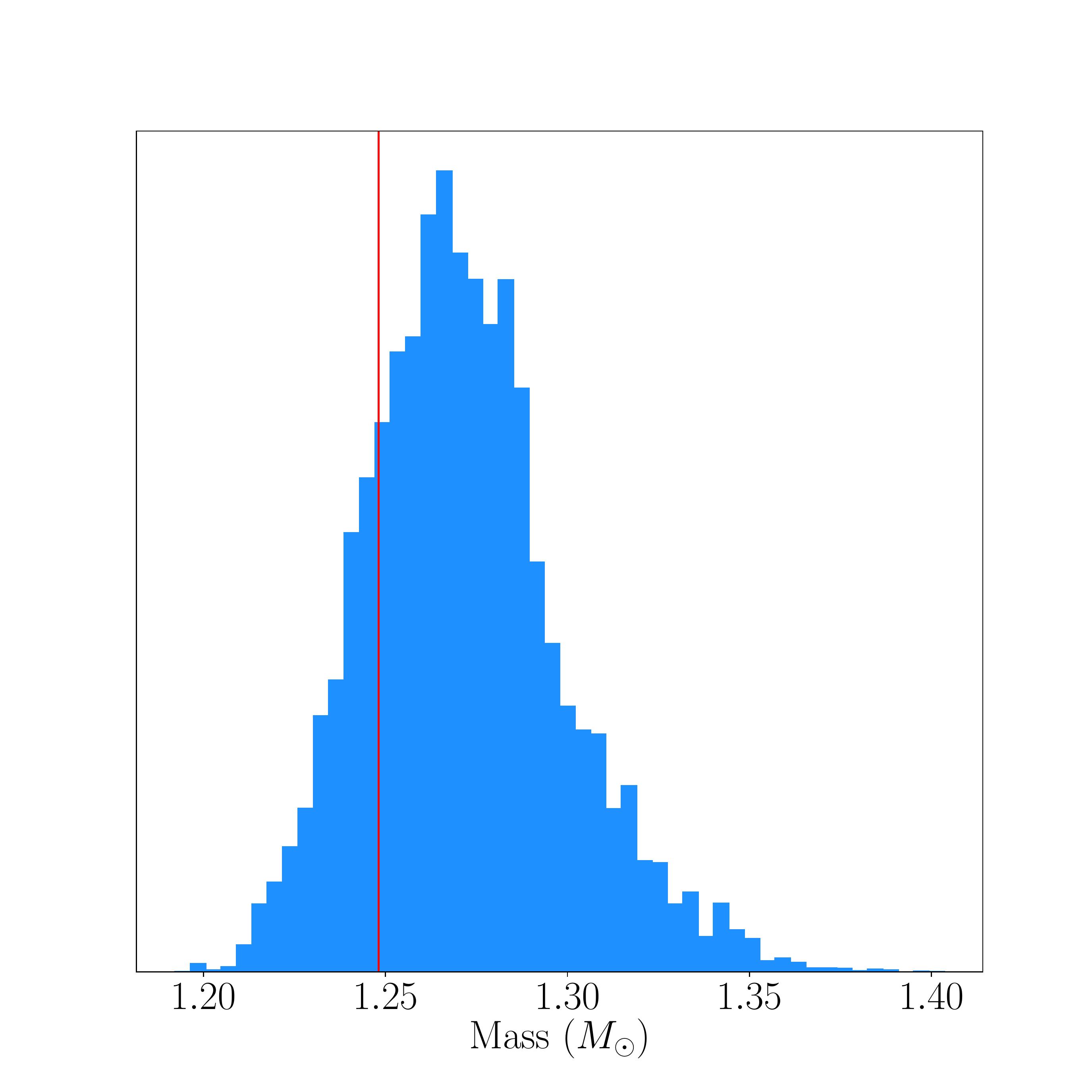}
 \end{minipage}	
 \end{flushleft}
	\caption{Probability distribution functions for the mass of $\rm{HD64121}$ (top left), $\rm{HD22532}$ (bottom), and $\rm{HD69123}$ (top right).The vertical red line in the plots indicates the position of the best model in the grid (without interpolation by AIMS).}
		\label{FigDistribMass}
\end{figure*}

\begin{figure*}
\begin{flushleft}
	\begin{minipage}{\textwidth}
  	\includegraphics[trim=70 25 70 95, clip, width=0.325\linewidth]{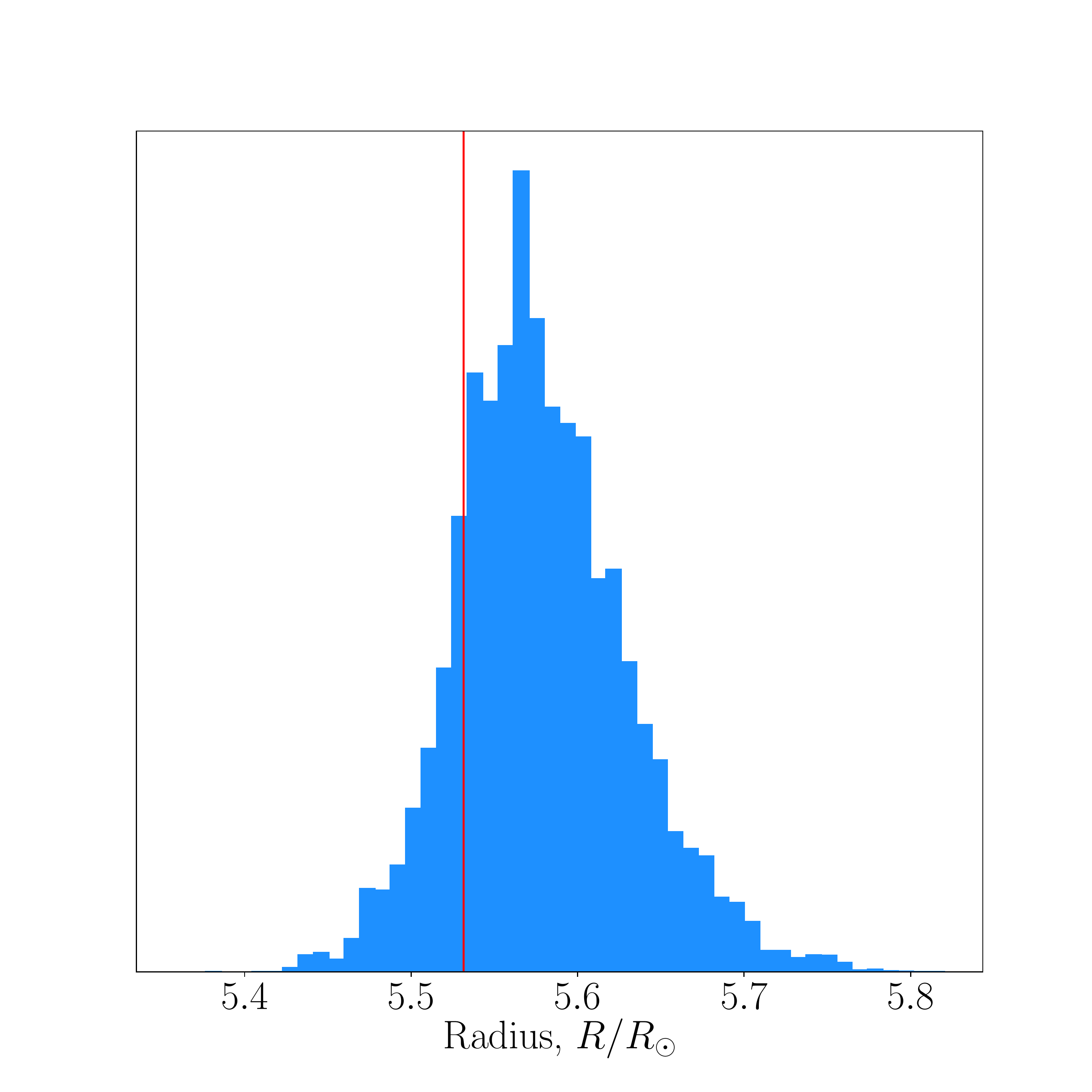} 
\includegraphics[trim=70 25 70 95, clip, width=0.325\linewidth]{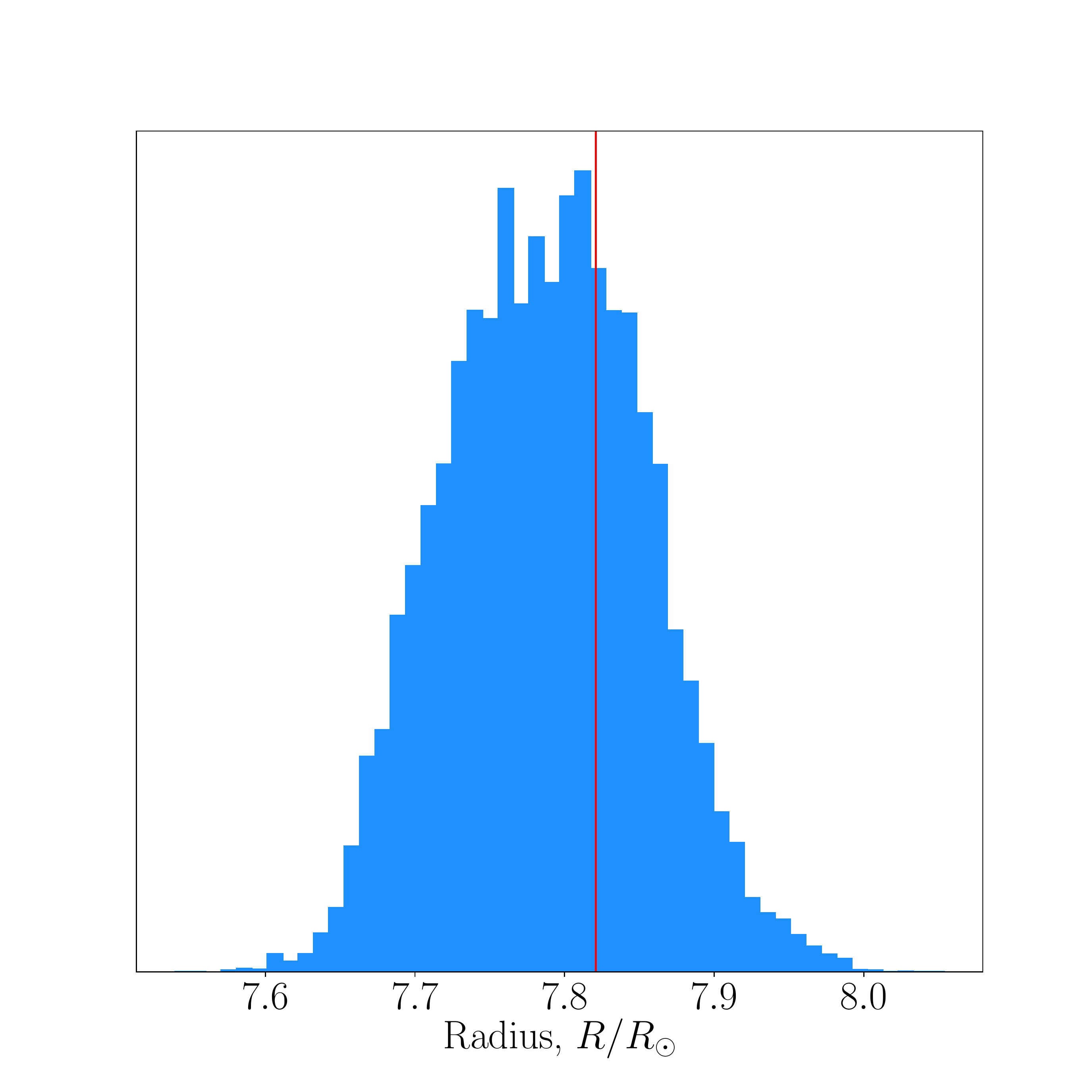}
 	\includegraphics[trim=70 25 70 95, clip, width=0.325\linewidth]{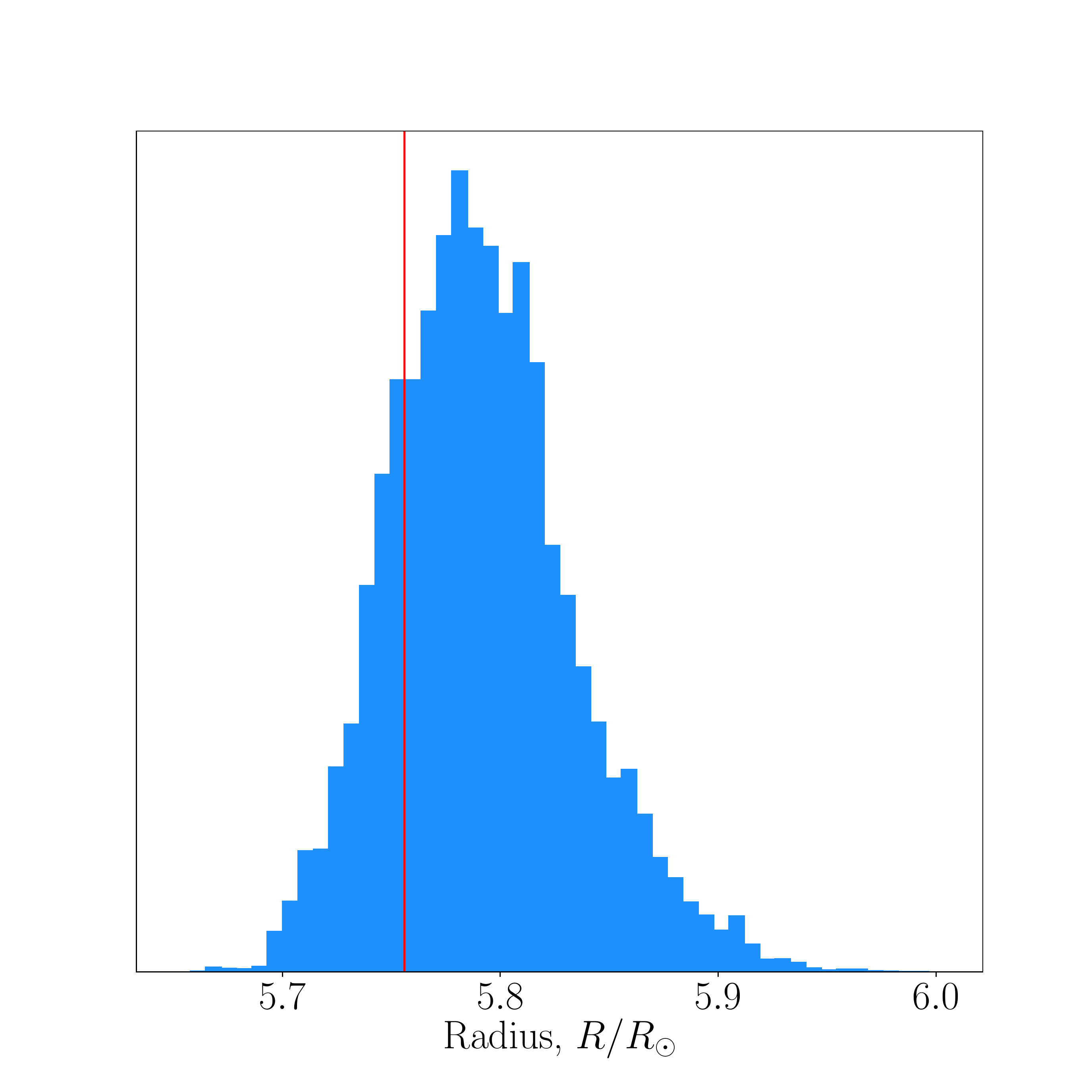}
 \end{minipage}	
 \end{flushleft}
	\caption{Probability distribution functions for the radius of $\rm{HD64121}$ (top left), $\rm{HD22532}$ (bottom), and $\rm{HD69123}$ (top right). The vertical red line in the plots indicates the position of the best model in the grid (without interpolation by AIMS).}
		\label{FigDistribRadius}
\end{figure*}

The modelling was carried out by fitting the individual radial modes of the stars, using the two-terms surface correction of \citet{Ball1}, in conjunction with $\rm{T_{eff}}$, $\left[ \rm{Fe/H}\right]$, and $\rm{L}$. As seen from Figs \ref{FigEchAIMS}, \ref{FigDistribMass} and \ref{FigDistribRadius}, the MCMC modelling has found a well-defined solution in all cases. The observed individual frequencies were corrected from the line-of-sight Doppler velocity shifts following the recommendations of \citet{Davies2014}.

However, we can see that the models have slightly different values from the radii from Ottoni et al. (submitted), which were not included in the dataset for the MCMC modelling. This can be regarded as a word of caution against the systematic use of individual frequencies in seismic modelling (as mentioned in \citet{Rendle2019} and in \citet{Buldgen2019}) as they heavily relied on the surface corrections used to determine the optimal solution \citep[see e.g.][for a discussion in the context of eclipsing binaries and stellar clusters]{Jorgensen2020}. If the individual modes are determined with very high precision, this can lead to extremely precise, although not fully accurate, solutions. In our specific cases, the seismic radius from the individual frequency modelling is only slightly different from the one determined from the parallaxes and spectroscopic observations, and within $1.5\sigma$ error bars.

Overall, the solution found by AIMS is of sufficient quality to serve as an initial condition for the next modelling step. However, since we did not exploit the information of the quadrupolar modes, we can assume that the modelling can be further refined. In addition, having clearly identified dipolar modes at our disposal would allow for an even more refined modelling, especially regarding core conditions (e.g. size of the helium core and boundary of the hydrogen burning shell). Unfortunately, as mentioned above, the dataset at our disposal was insufficient to fully identify them and thus exploit them in our theoretical modelling.

Indeed, the individual radial frequencies constrain the mean density almost solely, while their relatively high precision implies that the modelling focuses on reproducing them specifically\footnote{The issue would be even worse when directly fitting the individual frequencies of a main-sequence solar-like oscillator, due to higher relative precision and the larger number of constraints if non-radial modes are also included.}. Multiple strategies can be used to mitigate this aspect, one is to decrease the weight of the seismic constraints in the modelling or to increase the weight of the non-seismic constraints. None of these approaches are optimal in practice and the modelling strategies should be adapted to avoid such issues, especially if one wishes to study large samples automatically.

From a physical point of view, it should be noted that the apparent contradiction between seismic radii values and the radii determined from \textit{Gaia} parallaxes and spectroscopy may stem from the surface correction, or from the use of a fixed solar mixing-length parameter value for the grid\footnote{Other physical ingredients of the models could be at play but a detailed analysis of all degeneracies is beyond the scope of our study and would require a larger sample with higher-quality data.}.

\subsection{Seismic inversions and frequency differences}\label{SecFreqInv}

The inversions of the mean density were carried out using the reference models computed with AIMS, following the approach of \citet{ReeseDens}. As shown by \citet{Buldgen2019}, this approach is suitable to exploit the information of radial oscillations of first-ascent RGB and core Helium-burning stars, given that their evolutionary status is known. In the current study, this was the case for all our targets.

The inversion was computed following the variational formalism of \citet{Dziemboswki90}, using the SOLA method \citep{Pijpers} to carry out the inversion. The cost function minimised by the inversion is formally written as follows:
\begin{align}
\mathcal{J}_{\bar{\rho}}(c_{i})=&\int_{0}^{1}\left[K_{\mathrm{Avg}} - \mathcal{T}_{\bar{\rho}}\right]^{2}dx + \beta \int_{0}^{1}\left( K_{\mathrm{Cross}}\right)^{2}dx \nonumber \\ &+ \lambda \left[ 2 -\sum_{i}c_{i} \right] + \tan \theta \frac{\sum_{i}\left(c_{i}\sigma_{i}\right)^{2}}{<\sigma^{2}>}, \label{eq:CostSOLA}
\end{align}
where we have defined the target function of the inversion, the averaging and cross-term kernels,
\begin{align}
\mathcal{T}_{\bar{\rho}}&=4\pi x^{2} \frac{\rho}{\rho_{R}}, \\
K_{\mathrm{Avg}}&=\sum_{i}c_{i}K^{i}_{\rho,\Gamma_{1}}, \\
K_{\mathrm{Cross}}&=\sum_{i}c_{i}K^{i}_{\Gamma_{1},\rho},
\end{align}
with the radial position of an element of stellar plasma divided by the photospheric stellar radius $x=r/R$, being $\rho$ the local density, and $\rho_{R}=M/R^{3}$, with $M$ being the stellar mass and $R$ being the photospheric stellar radius. We have also introduced the parameters $\beta$ and $\theta$, defining the trade-off problem between the fit of the target, the contribution of the cross term and the amplification of observational error bars of the individual frequencies, denoted $\sigma_{i}$. The $K^{i}_{\rho,\Gamma_{1}}$ and $K^{i}_{\Gamma_{1},\rho}$ are the structural kernel functions, derived from the variational analysis of the pulsation equations and $<\sigma^{2}>=\frac{1}{N}\sum_{i=1}^{N}\sigma^{2}_{i}$ with $N$ being the number of observed oscillation modes. In Eq. \ref{eq:CostSOLA}, we have also defined the inversion coefficients $c_{i}$ and $\lambda$, a Lagrange multiplier. The third term is based on homologous reasoning described in \citet{ReeseDens}, which also serves to derive a non-linear generalisation of the method where the mean density, denoted $\bar{\rho}_{\mathrm{Inv}}$, was computed from
\begin{align}
\bar{\rho}_{\mathrm{Inv}}=\left(1+\frac{1}{2}\sum_{i}c_{i}\frac{\delta \nu_{i}}{\nu_{i}}\right)^{2}\bar{\rho}_{\mathrm{Ref}},
\end{align}
with $\bar{\rho}_{\mathrm{Ref}}$ being the mean density of the reference model and $\frac{\delta \nu_{i}}{\nu_{i}}$ being the relative differences between the observed and theoretical frequencies defined as $\frac{\nu_{\mathrm{Obs}}-\nu_{\mathrm{Ref}}}{\nu_{\mathrm{Ref}}}$. If we use the non-linear generalisation, the errors on the mean density value from the inversion are given by
\begin{align}
\sigma_{\bar{\rho}_{\mathrm{Inv}}}=\bar{\rho}_{\mathrm{Ref}}\left(1+\frac{1}{2}\sum_{i}c_{i}\frac{\delta \nu_{i}}{\nu_{i}} \right)\sqrt{\sum_{i}c^{2}_{i}\sigma^{2}_{i}}.
\end{align}

The results of the inversion for the mean density are the following: $\bar{\rho}^{\rm{HD69123}}_{\rm{Inv}}=0.00436 \pm 0.00004$ $\rm{g/cm^{3}}$, $\bar{\rho}^{\rm{HD22532}}_{\rm{Inv}}=0.00907 \pm 0.00007$ $\rm{g/cm^{3}}$, and $\bar{\rho}^{\rm{HD64121}}_{\rm{Inv}}=0.01025 \pm 0.00025$ $\rm{g/cm^{3}}$. The uncertainties were determined from both the analysis of model-dependencies as well as the use of various surface corrections, as was done in \citet{Buldgen2019}.

The mean density determined from the inversion was then used as a constraint for an additional forward modelling step combined with the small frequency separations. This additional step was carried out with a Levenberg-Marquardt minimisation technique, which allowed us to consider additional free parameters and to compute the evolutionary models directly with CLES. 

The constraints used for this local minimisation were as follows: $\left[\rm{Fe}/\rm{H} \right]$, $\rm{T_{eff}}$, $\rm{L}$, $\bar{\rho}$, and the individual $d_{0,2}$. The free parameters were the mass, the age, the initial chemical composition, and the mixing-length parameter of convection. The main motivation for the use of the individual $d_{0,2}$ as constraints for the modelling is that \citet{Montalban2010, Montalban2012} have shown that these quantities are sensitive to the stellar mass for a given value of the large frequency separation due to the radius sensitivity of $d_{0,2}$ for a given mean density.  

\begin{figure}
	\centering
		\includegraphics[width=8cm]{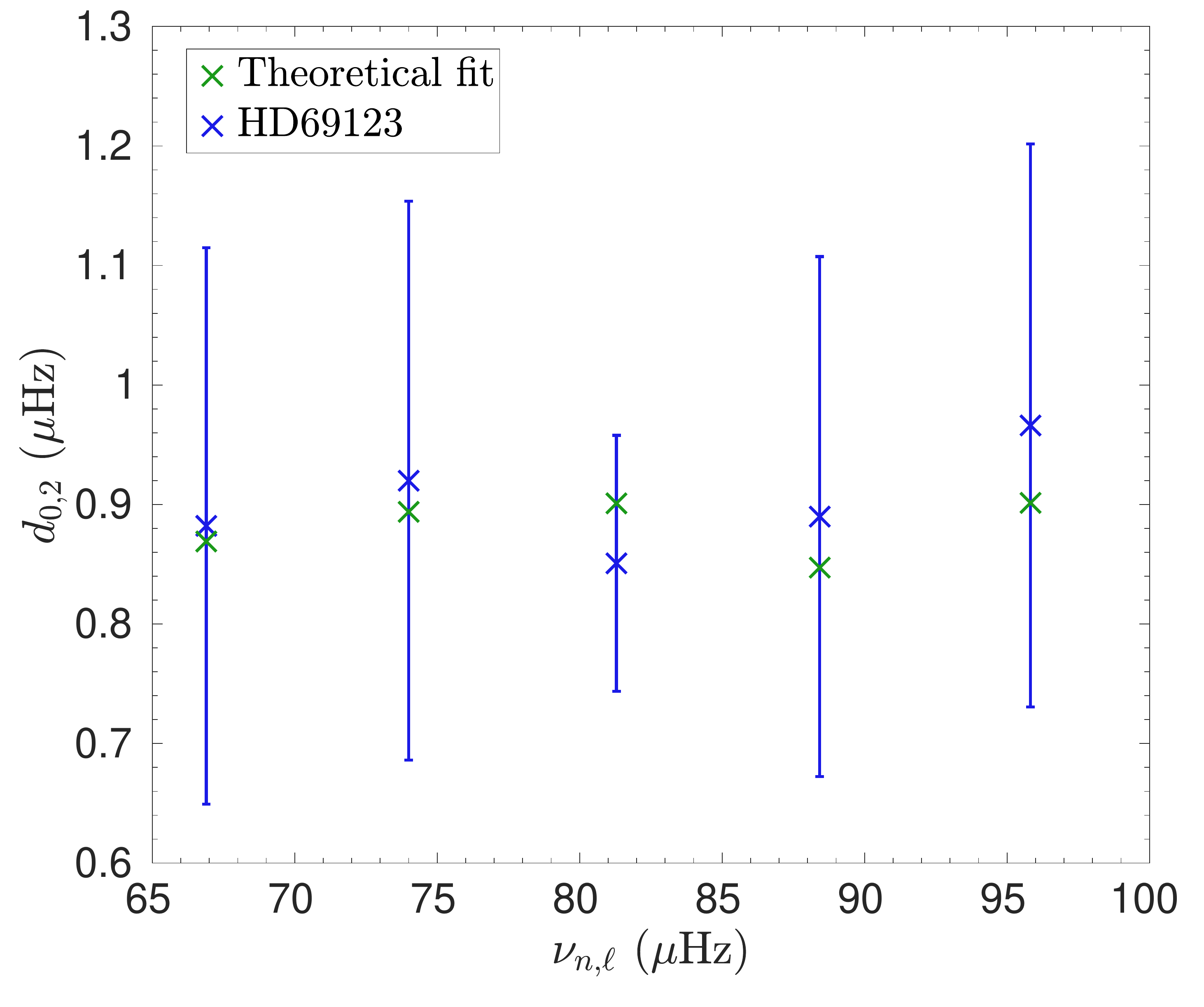}
	\caption{Comparisons between the theoretical and the observed $d_{0,2}$ for TIC$146264536$/HD$69123$.}
		\label{Figd02}
\end{figure}

\begin{figure*}
	\centering
		\includegraphics[width=18cm]{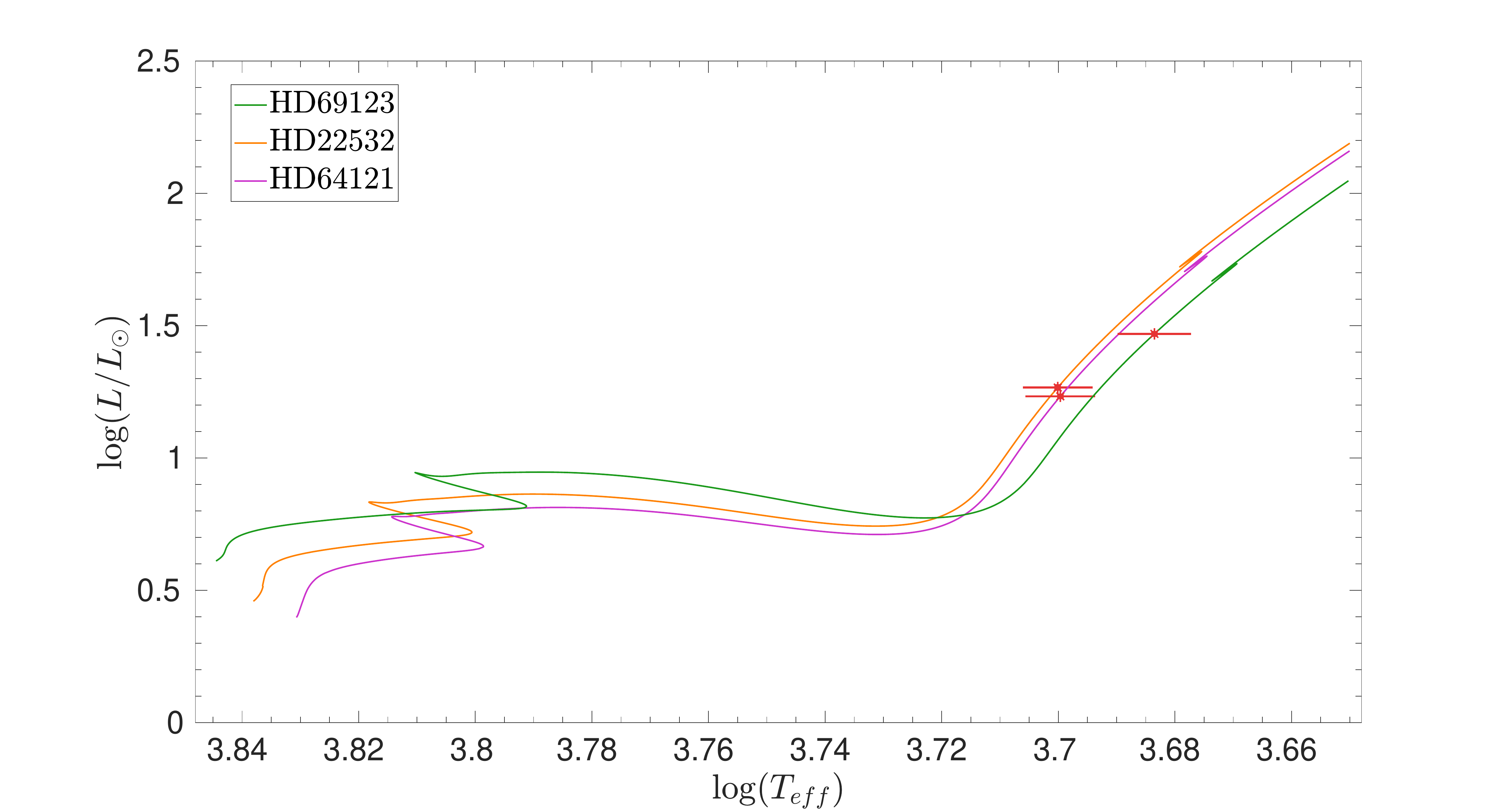}
	\caption{Hertzsprung-Russel diagram showing the tracks of all targets, the location of the current position of the system being given by the red star.}
		\label{FigHRAll}
\end{figure*}

The results of this third modelling step are given in table~\ref{tabFinMod} and the agreement in terms of an individual small separation are illustrated in Fig. \ref{Figd02} for one target of our sample. In Fig. \ref{FigHRAll}, we illustrate the evolutionary tracks computed with CLES for each target, using the optimal parameters derived from the seismic modelling. 

Given this third modelling step, we can see that a better agreement in terms of the radius was found for the modelling, implying a better mass determination. However, given the degenerate interplay between the initial chemical composition and the mixing-length parameter that impacts our final results, we can safely state that the estimation of the age of the targets could still be refined by exploiting the information of individual dipolar mixed modes, following for example the approach of \citet{Deheuvels2011} and \citet{Noll2021}. In our case, however, these modes were not identified in the power spectrum and were thus absent from the modelling. 

\begin{table*}[t]
\caption{Parameters of the optimal stellar models found from the Levenberg-Marquardt minimisation.}
\label{tabFinMod}
  \centering
\begin{tabular}{r | c | c | c | c }
\hline \hline
\textbf{Identifiers}& $\rm{HD22532}$ & $\rm{HD64121}$ & $\rm{HD69123}$\\ \hline
$\rm{M}$ $\rm{(M_{\odot})}$  &$1.23\pm0.07$&$1.19\pm0.07$&$1.43\pm0.06$\\
$\rm{R}$ $\rm{(R_{\odot})}$ &$5.77\pm0.04$&$5.49\pm0.04$&$7.78\pm0.05$\\
$\rm{X_{0}}$ &$0.71\pm0.02$&$0.72\pm0.02$&$0.72\pm0.01$\\
$\rm{Z_{0}}$ &$0.0081\pm0.001$&$0.0078\pm0.001$&$0.0137\pm0.001$\\
$\rm{\alpha_{MLT}}$ &$2.06\pm0.1$& $2.15\pm0.1$&$2.08\pm0.1$\\
$\rm{Age}$ $\rm{(Gyr)}$ &$4.03\pm1.1$&$4.51\pm1.0$&$3.31\pm1.5$\\
\hline
\end{tabular}

\end{table*}

It is also clear that the values in tab. \ref{tabFinMod} are subject to the intrinsic limitations of the stellar evolution code and could substantially vary if we changed the physics of the models on both the micro- and macroscopical level. However, for each star, we could also provide a model independent mass interval from the combination of the mean density determined by the inversion and the radius determined from the \textit{Gaia} parallaxes and the spectroscopic data. These intervals are provided in table~\ref{tabMassModInd}. While these values were computed independently from any stellar model, they still suffer from some limitations, namely the intrinsic accuracy of the inversion techniques discussed in \citet{Buldgen2019} as well as that of the radius determination. In other words, these mass intervals are sensitive to the stellar spectroscopic and astrometric observations, the bolometric correction and the extinction laws used in conjunction with the \textit{Gaia} data.

\begin{table*}[t]
\caption{Model-independent mass intervals from seismic inversions and radii from \textit{Gaia} parallaxes and spectroscopy.}
\label{tabMassModInd}
  \centering
\begin{tabular}{r | c | c | c | c | c}
\hline \hline
\textbf{Identifiers}& $\rm{HD22532}$ & $\rm{HD64121}$  & $\rm{HD69123}$\\ \hline
$\rm{M}$ $\rm{(M_{\odot})}$ &$1.21\pm0.10$&$1.18\pm0.11$&$1.43\pm0.12$\\
$\rm{R}$ $\rm{(R_{\odot})}$ &$5.70\pm0.12$&$5.44\pm0.11$&$7.72\pm0.17$\\
\hline
\end{tabular}

\end{table*}

When compared with the values determined by both AIMS and the seismic scaling relations, we can see that the match is quite good for all three targets. However, at the desired level of precision, it appears that the disagreements can be of more than $1\sigma$ for some targets, as a result of the correction factors. Again this illustrates the lack of robustness of the scaling relations at this level of precision and the need for a detailed stellar modelling procedure to study planetary systems in detail. 

\section{Orbital evolution}\label{SecOrbital}

In this section, we aim to study the orbital evolution of the three planetary systems HD22532 (TIC200841704), HD64121 (TIC264770836), and HD69123 (TIC146264536). The orbital parameters used in this study are presented in Table~\ref{param_orb}, in which we report the orbital period in units of days, the values of the semi-major axis in AU, and finally the upper limits on the planetary masses expressed in units of Jupiter masses. We note that the inclinations of the systems are unknown.

First, we computed stellar models representative of the host stars of the systems with the CLES stellar evolution code. We used the stellar parameters derived from the asteroseismic modelling as input and constraints. Second, we coupled the stellar models to our orbital evolution code \citep{Privitera2016A, Privitera2016B, Privitera2016c, Meynet2017, Rao2018}, taking the exchange of angular momentum between the planetary orbit and the star into account, occurring through the dissipation of tides in the stellar convective envelope. Thanks to this approach, we can investigate whether dynamical and/or equilibrium tides had a significant impact during the past history of the systems.

The physics adopted in the orbital evolution code is described in more detail in the work of \citet{Rao2018}. Here, we briefly recall the fundamental equations. The equation for the total change of the orbital distance reads as follows:

\begin{align}
\rm \left(\dot{a}/a \right) = - \dfrac{\dot{M}_{\star} +\dot{M}_{pl}}{ M_{\star} + M_{pl}} - \dfrac{2}{M_{pl} v_{pl}} \left[ F_{fri} + F_{gra} \right] + \left( \dot{a}/a\right)_{t}, \label{eq:total_a}
\end{align}

where the term $\rm \dot{M}_{\star} = - \dot{M}_{loss}$ indicates the stellar mass loss rate, $\rm M_{pl}$ is the mass of the planet, $\rm \dot{M}_{pl}$ is the rate of change of the planetary mass and $\rm v_{pl}$ is the planetary orbital velocity. Futhermore, $\rm F_{fri}$ and $\rm F_{gra}$ represent the contribution due to the frictional and gravitational drag forces respectively, whose expressions are taken as in \citet{Villaver2009, Mustill2012, Villaver2014}. The term $\rm \left( \dot{a}/a \right)_{t} $ includes the impacts of dynamical and equilibrium tides. In particular, the expression for equilibrium tides \citep{Zahn1966, Alexander1976, Zahn1977,  Zahn1989, LS1984b, Villaver2009, MV2012, Villaver2014} is taken as in \citet{Privitera2016B} and their contribution is accounted for only when a stellar convective envelope is present. The equation for equilibrium tides is
 
\begin{align}
\rm \left( \dot{a}/a \right)_{eq} = \frac{f}{\tau_{cz}} \frac{M_{env}}{M_{\star}} q(1+q) \left( \frac{R_{\star}}{a} \right)^8 \left( \frac{\Omega_{\star}}{\omega_{pl}} - 1\right).\label{eq:equi_tides}
\end{align}

The term $\rm f$ is a numerical factor obtained from integrating the viscous dissipation of the tidal energy across the convective zone \citep{Villaver2009}, $\rm M_{env}$ is the mass of the convective envelope, $\rm q$ is the ratio between the mass of the planet and the one of the star ($\rm q = M_{pl}/M_{\star}$), $\rm \Omega_{\star}$ is the stellar surface rotation rate, $\rm \omega_{pl} = 2\pi/ P_{orb}$ is the orbital frequency of the planet, and $\rm \tau_{cz} $ is the convective turnover timescale.

As mentioned above, we also accounted for the impact of dynamical tides, in the form of a frequency-averaged tidal dissipation of inertial waves excited in the convective envelope of a rotating star, for which the Coriolis force is the restoring force \citep{BolmontMathis2016,Gallet2017,Bolmont2017,Benbakoura2019}. In our study, we assume the planet is on a circular-coplanar orbit around its host star. In this context, the effect of dynamical tides is accounted for whenever the condition $\rm \omega_{pl} < 2~\Omega_{\star}$ is satisfied since a planet orbiting a star on a circular-coplanar orbit is able to excite inertial waves when the orbital frequency is lower than twice the stellar rotation rate \citep{Ogilvie2007}. The expression of the change in orbital distance under the effects of dynamical tides is taken from \citet{Ogilvie2013} and \citet{Mathis2015} and is written as

\begin{align}
\rm \left( \dot{a}/a \right)_{dyn} = \left( \dfrac{9}{2Q^{\prime}_{d}} \right)q \omega_{pl} \left( \frac{R_{\star}}{a} \right)^5 \dfrac{(\Omega_{\star} - \omega_{pl})}{\mid \Omega_{\star} - \omega_{pl}\mid},
\label{eq:dyn_tides}
\end{align}

with $\rm Q^{\prime}_{d} = 3/(2D_{\omega})$ and  $\rm D_{\omega} = D_{0\omega}D_{1\omega}D_{2\omega}^{-2}$. The `D' terms are defined as

\begin{align}
\begin{cases}
\rm D_{0\omega} = \dfrac{100\pi}{63} \epsilon^{2} \dfrac{\alpha^5}{1 - \alpha^5} (1 - \gamma)^2,\\
\rm D_{1\omega} = (1 - \alpha)^4 \left( 1 + 2\alpha + 3\alpha^2 + \frac{3}{2} \alpha^3 \right)^2 ,\\
\rm D_{2\omega} = 1 + \frac{3}{2}\gamma + \frac{5}{2 \gamma}\left( 1 + \frac{\gamma}{2} - \frac{3 \gamma^2}{2} \right) \alpha^3 - \frac{9}{4}\left(1 - \gamma\right)\alpha^5  ,
\end{cases}
\end{align}

where $\rm \alpha = R_{c}/R_{\star}$, $\rm \beta = M_{c}/M_{\star}$, $\rm \gamma = \dfrac{\alpha^3 (1 - \beta)}{\beta (1 - \alpha^3)}$, $\rm \epsilon = \dfrac{\Omega_{\star}}{\sqrt{\dfrac{\rm GM_{\star}}{\rm R_{\star}^3}}}$. Furthermore, $\rm M_{c}$ and $\rm R_{c}$ represent the mass and the radius of the radiative core. The term $\rm D_{\omega}$ was computed in \citet{Ogilvie2013} as the frequency-averaged tidal dissipation. 

Whenever the orbital distance of the planet becomes equal to the corotation radius, defined as the distance at which $\rm \omega_{pl} = \Omega_{\star}$, tides become inefficient. When this condition is not satisfied, the tides widen or shrink the orbit, when the planet is beyond or inside the corotation radius, respectively.

Magnetic braking at the stellar surface is taken into account by using the formalism of \citet{Matt2015,Matt2019}, for which the magnetic torque writes
\begin{align}
\rm \dfrac{dJ}{dt} =
\begin{cases}
\rm -T_{\odot} \left(\dfrac{R}{R_{\odot}} \right)^{3.1} \left( \dfrac{M}{M_{\odot}} \right)^{0.5} \left(\dfrac{\tau_{cz}}{\tau_{cz \odot}} \right)^{p} \left(  \rm \dfrac{\Omega}{\Omega_{\odot}} \right)^{p+1} &, \rm \text{if} ~ \left( Ro > Ro_{\odot}/\chi \right),\\
\rm -T_{\odot} \left(\dfrac{R}{R_{\odot}} \right)^{3.1} \left( \dfrac{M}{M_{\odot}} \right)^{0.5} \chi^{p} \left( \dfrac{\Omega}{\Omega_{\odot}} \right) & , \rm \text{if} ~ \left( Ro \leq Ro_{\odot}/\chi \right) ,
\end{cases}
\end{align}

where $\rm R_{\odot}$ and $\rm M_{\odot}$ are the radius and the mass of the Sun, and $\rm R$ and $\rm M$ are the radius and the mass of the stellar model. Furthermore, $\rm \tau_{cz}$ is the convective turnover timescale and $\rm Ro$ is the Rossby number, defined as the ratio between the stellar rotational period and the convective turnover timescale ($\rm Ro = P_{\star}/\tau_{cz}$). The quantity $\rm \chi \equiv Ro_{\odot} / Ro_{sat}$ indicates the critical rotation rate for stars with a given $\rm \tau_{cz}/\tau_{cz_{\odot}}$, defining the transition from saturated to unsaturated regime. We considered $\rm \chi$ to be equal to $\rm 10$ as in \citet{Matt2015} and \citet{Eggenberger2019a}. The exponent $\rm p$ was considered equal to $\rm 2.3$ and the constant $\rm T_{\odot}$ was calibrated in order to reproduce the solar surface rotation rate \citep{Eggenberger2019a}.

We accounted for the evaporation of the planetary atmosphere following the formalism of the Jeans escape regime \citep{Villaver2007} or the one of the hydrodynamic escape regime, depending on the properties of the planetary system under study. When using the hydrodynamic escape regime, we computed the planetary mass loss by using the energy limited formula as in the work of \citep{Lecavelier2007,Erkaev2007}.

For each of the systems considered, we initially studied the evolution of the orbits under the impact of dynamical and equilibrium tides at a fixed planetary mass. Therefore, we did not account for the atmospheric mass loss at this stage. We took the values of the semi-major axis and the masses reported in Table~\ref{param_orb} as initial input and tested whether we were able to reproduce the position of the planets at the ages of the systems.

\begin{figure}
\centering
\includegraphics[width=0.47\textwidth]{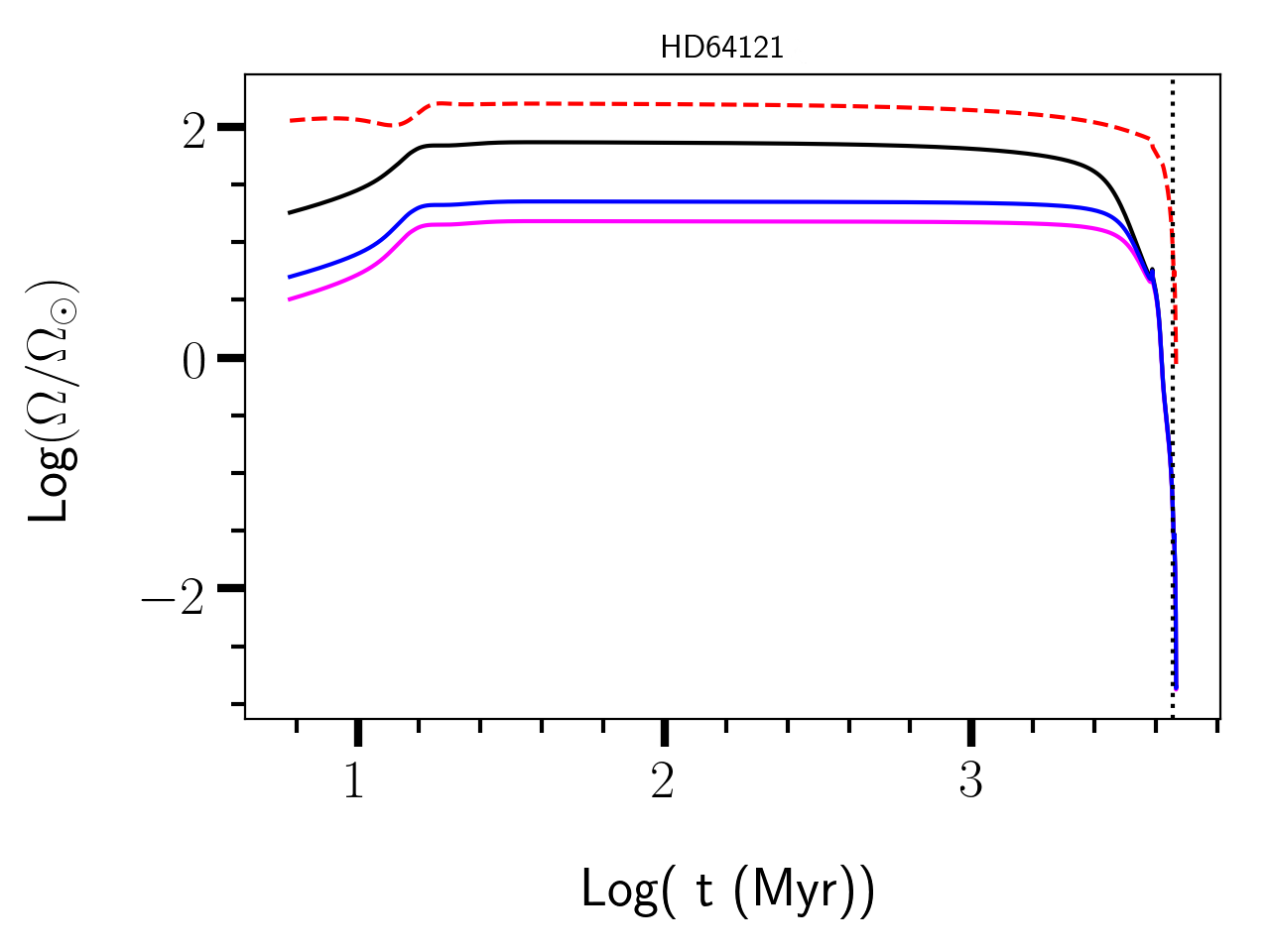}
\caption{Surface angular velocity evolution for HD64121 (TIC 264770836) in the case of 	 fast rotator ($\rm \Omega_{in} = 18 \times \Omega_{\odot}$, black solid line), a medium rotator ($\rm \Omega_{in} = 5 \times \Omega_{\odot}$, blue solid line) and a slow rotator ($\rm \Omega_{in} = 3.2 \times \Omega_{\odot}$). The red dashed line represents the critical velocity limit. The black dotted vertical line indicates the age of the system.}
\label{Om_Surf}
\end{figure}

The rotational history of the star being unknown, we shall considered a range of initial surface rotation rates representative of different kinds of rotators. We considered three different values for the initial surface rotation rate (namely $\rm \Omega_{in} = 3.2, 5$, and $\rm 18~ \Omega_{\odot}$), covering the range for slow, medium and fast rotators as deduced from surface rotation rates of solar-type stars observed in open clusters at different ages \citep{Eggenberger2019a}. A disk lifetime of 2 Myr and 6 Myr was used for the fast and medium-slow rotating case, respectively. During the disk-locking phase, we assumed the surface angular rotation remained constant. We also assumed the star rotating as a solid body. In Fig.~\ref{Om_Surf}, we report the evolution of the surface rotation rates computed for HD64121. The tracks show that at the beginning of the evolution the star experiences a spin-up due to the rapid contraction of the structure occurring during the PMS phase, reaching a maximum at the age of $\rm \sim 16$ Myr. After the peak, the braking due to magnetised stellar winds started to be efficient. Nevertheless, we notice that for a star such as HD64121, the efficiency of this process is reduced by the presence of a very shallow external convective envelope. The rotation rate evolutionary tracks therefore remain quite flat for the duration of the MS phase, until the star reaches the RGB, expanding and significantly reducing its rotation rate.

It is worth noting that when considering a value of $\rm \Omega_{in}$ as large as $\rm 18~\Omega_{\odot}$ and a disk-locking timescale of 2 Myr, the surface rotation rate of the star rapidly reaches overcritical values. We attempted to mitigate this effect by using a longer disk-locking timescale of 6 Myr as in the case of the slow and medium rotators, and we managed to maintain the track at subcritical values (as shown in Fig.~\ref{Om_Surf}). This method did not work for the systems HD22532 and HD69123, for which we excluded the fast rotating track as a viable rotational history of the host star. This result emphasizes how such an analysis of planetary systems can help to lift the degeneracy on the possible rotational histories of the stars. 

\begin{figure}
\centering
        \includegraphics[width=0.45\textwidth]{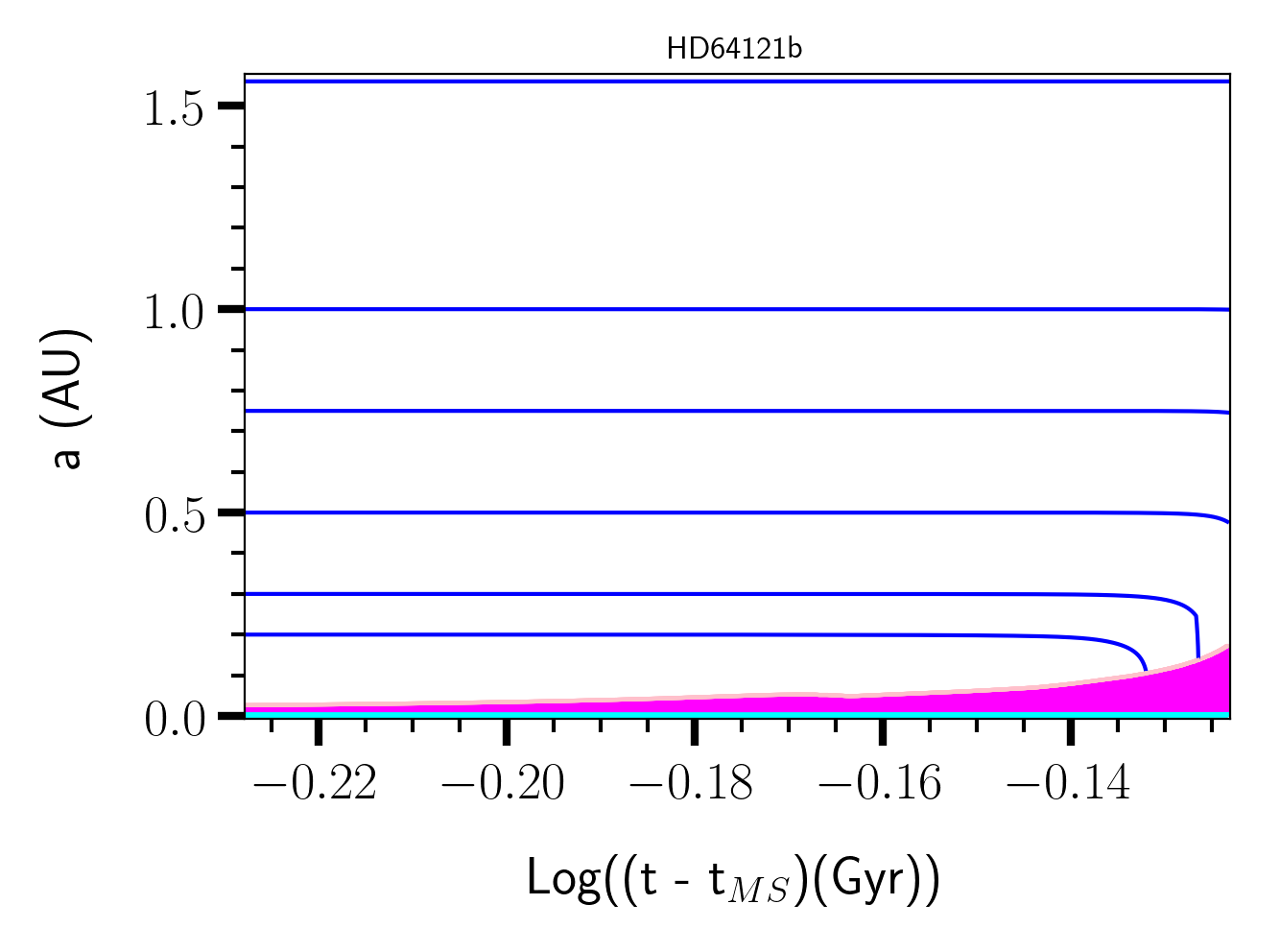}    
\caption{Impact of a change in orbital distance (blue solid lines) for HD64121b, for a host-star initial surface rotation of $\rm \Omega_{in} = 5 \times \Omega_{\odot}$. The magenta area shows extension of the stellar convective envelope, while the cyan area represents the extension of the stellar core.}
\label{Orbit_Evol}
\end{figure}

Using this initial set up, we did not find any appreciable evolution of the orbital distances for our systems. This is an expected outcome, which is in agreement with previous results reported in \citet{Privitera2016B}, given the relatively large value for the orbital distances together with the moderate planetary masses. Even when considering the highest value for the initial surface rotation for HD64121, we did not obtain significant impacts on the planetary orbit. We explored a range of lower initial orbital distances for HD64121b, using an intermediate value for the surface rotation rate of the host star ($\rm \Omega_{in} = 5~\Omega_{\odot}$), in order to determine the maximal value below which the orbit of the planet would be significantly affected by tides, eventually leading to an engulfment at earlier times with respect to the age of the system. As shown in the top panel of Fig.~\ref{Orbit_Evol}, we find that the maximal value of the initial orbital distance below which the orbit would be impacted and the planet subsequently engulfed is $\rm a_{max} \approx 0.5$ AU. We would expect $\rm a_{max}$ to shift at higher values when considering larger surface rotation rates ($\rm \Omega_{in} = 18~\Omega_{\odot}$).

Subsequently, we computed the evolution of the systems by also including the impact of atmospheric evaporation. For the three systems considered, we followed the evolution of the escape parameter, assuming a value of the Bond albedo of $\rm A = 0.5$ \citep{Villaver2007}. We find that the escape parameters always have values above the threshold representing the switch from Jeans escape to hydrodynamic escape regime, here considered as $\rm E \gtrsim 20$ \citep{Villaver2007, Villaver2009}. In this context, the evaporation rates computed within the Jeans escape regime conditions resulted in a negligible impact on the evolution of the planetary masses.\\

\begin{table*}
\caption{Orbital parameters of HD22532b, HD64121b, and HD69123b.}
\label{param_orb}
  \centering
\begin{tabular}{r | c | c | c  }
\hline \hline
\textbf{Identifiers}& $\rm{HD22532b}$ & $\rm{HD64121b}$ & $\rm{HD69123b}$ \\ \hline
$\rm P ~(days)$ & 872.6~$\pm$~2.8 & 623.0~$\pm$~3.4 & 1193.3~$\pm$~7.0\\
$\rm a ~(AU)$ &$1.900 \pm 0.004$ & $1.510 \pm 0.006$ & $2.482 \pm 0.010$\\
$\rm  M_{pl} sin(i) ~ (M_{J})$ & $2.12 \pm 0.09$ & $ 2.56 \pm 0.19$ & $ 3.04 \pm 0.16$\\
\hline
\end{tabular} \\
\end{table*}

\section{Conclusion}\label{SecConc}

In this study, we have carried out a detailed comparative modelling of three exoplanet-host red giant stars. Each star is orbited by a long-period Jupiter-like planet that has been detected in radial velocities within the CASCADES survey (Ottoni et al. submitted) using the CORALIE spectrograph. We used the TESS photometric time series of each target to obtain seismic constraints. We extracted the individual radial and quadrupolar oscillation frequencies using the \texttt{PBjam} peakbagging package \citep{Nielsen2020} as well as the global seismic parameters $\left\langle  \Delta \nu \right\rangle $ and $\nu_{\rm{max}}$, following the approach of \citet{Elsworth_2020}. Then we have analysed various approaches to determine the stellar fundamental parameters and computed the dynamical evolution of the system as well as the evaporation of the planetary atmospheres for each target, using the optimal seismic solution for the stellar properties. 

The first step was to determine the stellar properties from the seismic scaling relations. As expected, a correct order of magnitude guess is obtained for this case, but the main issue is the strong dependency on the underlying correction of the scaling relations, which can lead to inaccurate results. We conclude that for the purpose of the detailed analysis carried out here, the seismic scaling relations do not offer a sufficient degree of accuracy and robustness. However, this does not imply that they cannot be useful for the analysis of large samples, for which extensive computations are either not justified or not accessible due to limitations of the seismic data. In that sense, using global seismic parameters such as $\left\langle  \Delta \nu \right\rangle $ and $\nu_{\rm{max}}$, either in conjunction with model grids or with the scaling relations, still offers a uniform approach to do comparisons on large samples \citep[such as done for example by][for the case of the  \textit{Gaia} offset analysis]{Khan2019}. 

In a second modelling step, we carried out a fit of the individual radial modes in conjunction with non-seismic constraints using the AIMS software and a predetermined grid of stellar models. We have shown that these results are already more accurate. From this second step, we could determine a value for the mean density of each star by means of the SOLA inversion \citep{ReeseDens}, which could then be used both as an additional constraint and to derive a model-independent mass value when combined with \textit{Gaia} parallaxes and spectroscopic constraints. Such a constraint is of great interest for both the purpose of exoplanetology and also for even more advanced modelling aiming to exploit the information of dipolar oscillation modes. 

Finally, the last step of our seismic modelling was carried out using a local minimisation algorithm and computing the stellar evolution models on the fly, fitting a series of seismic and non-seismic constraints. This allowed us to refine the fundamental properties of the star and to eliminate any remaining inconsistencies regarding the seismic and parallax-estimated radii for all targets. All the masses were consistent with the model-independent interval derived above, as a result of the constraints used in the minimisation. The evolutionary tracks found from this procedure were then used to study the dynamical evolution and the evaporation of the planetary atmospheres for each planet around the stars of our sample. 

From this analysis, we have shown that no significant migration of the planets is expected, given their long periods. We have also shown that two out of three stars could not have had initial rotation velocities of $\rm 18~ \Omega_{\odot}$, as they would have then reached the critical velocity. 

To summarise, we have demonstrated the importance of a thorough seismic modelling for the purpose of exoplanetology and illustrated its application in studying the fate of three long period planets around red giant stars. It is, however, evident that the modelling technique can be even more thorough than what is shown here should the data quality improve or additional strategies be developed in the future. In both cases, the conclusion that a refined approach is required to study the properties of exoplanetary systems in depth would still hold. 

\section*{Acknowledgements}

We thank the referees for their suggestions and careful reading of the manuscript. G.B. acknowledges fundings from the SNF AMBIZIONE grant No. 185805 (Seismic inversions and modelling of transport processes in stars). C.P. acknowledges fundings from the Swiss National Science Foundation (project Interacting Stars, number 200020-172505). P.E. has received funding from the European Research Council (ERC) under the European Union's Horizon 2020 research and innovation programme (grant agreement No 833925, project STAREX). WHB, GD and AL thanks the UK Science and Technology Facilities Council (STFC) for support under grant ST/R0023297/1. This work has received funding from the European Research Council (ERC) under the European Union’s Horizon 2020 research and innovation programme (CartographY GA. 804752). The research leading to this paper has received funding from the European Research Council (ERC grant  agreement No. 772293 for the project ASTEROCHRONOMETRY). This article used an adapted version of InversionKit, a software developed within the HELAS and SPACEINN networks, funded by the European Commissions's Sixth and Seventh Framework Programmes. 

\bibliography{biblioarticleTess2}

\appendix
\section{Additional figures}\label{SecAddPlots}

\begin{figure}
    \centering
    \begin{subfigure}[t]{0.47\textwidth}
        \centering
        \includegraphics[width=1.0\textwidth]{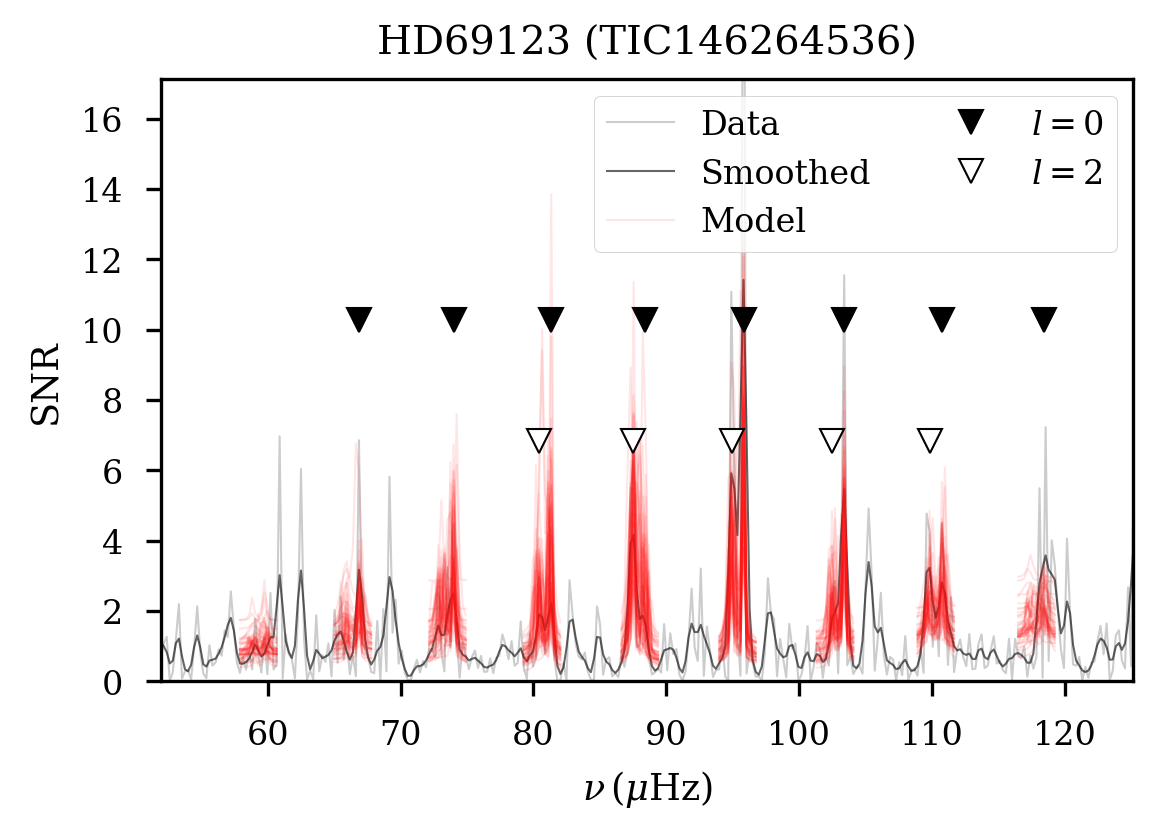}
    \end{subfigure}\\
    ~
    \begin{subfigure}[t]{0.47\textwidth}
        \centering
        \includegraphics[width=1.0\textwidth]{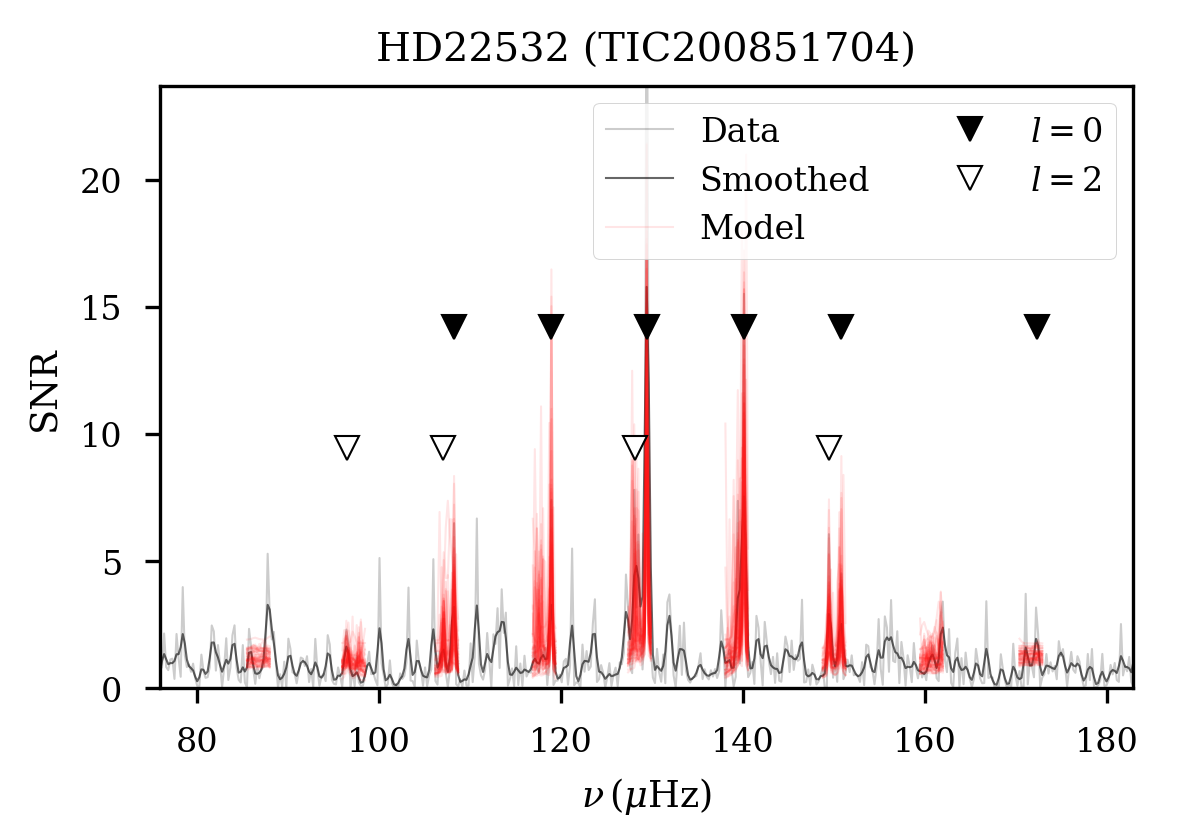}
    \end{subfigure}\\
    ~
    \begin{subfigure}[t]{0.47\textwidth}
        \centering
        \includegraphics[width=1.0\textwidth]{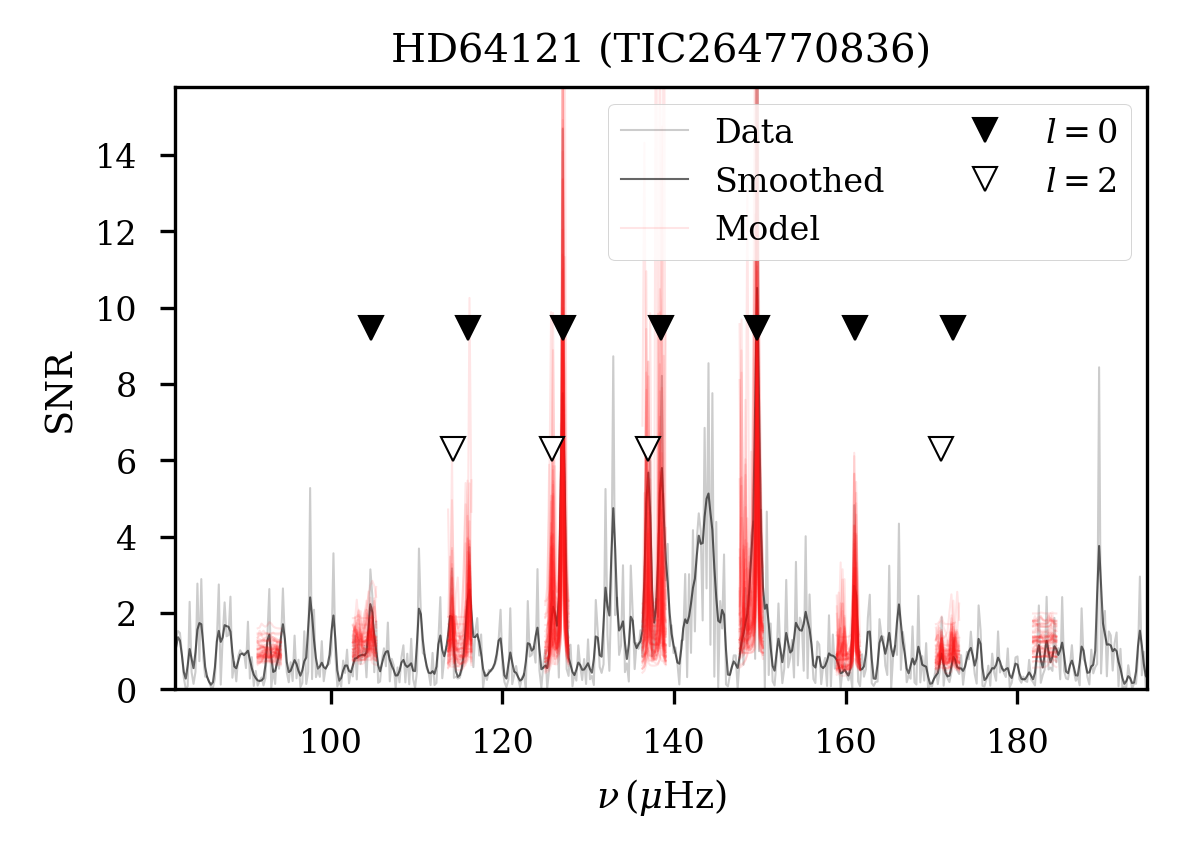}
    \end{subfigure}
    \caption{Observational power spectrum for each star. Locations of the radial ($\ell=0$) and quadrupolar ($\ell=2$) oscillation modes are given by white and black triangles respectively.}
    \label{fig:seismo_PS}
\end{figure}

\end{document}